\pdfoutput=1

\documentclass[12pt]{article}

\usepackage{amsfonts}
\usepackage{amsmath}
\usepackage{amssymb}
\usepackage{bigints}
\usepackage{booktabs}
\usepackage[nosort]{cite}
\usepackage{color}
\usepackage{dsfont}
\usepackage{float}
\usepackage{framed}
\usepackage{graphicx}
\usepackage{indentfirst}
\usepackage{mathrsfs}
\usepackage{multirow}
\usepackage{pdflscape}
\usepackage{setspace}
\usepackage{subdepth}
\usepackage{subfig}
\usepackage{titlesec}
\usepackage[dotinlabels]{titletoc}
\usepackage{wrapfig}
\usepackage[all]{xy}
\usepackage{young}
\usepackage[vcentermath]{youngtab}
\usepackage{relsize}
\usepackage{stackengine}
\usepackage{datetime}
\usepackage{physics}

\usepackage{hyperref}

\usepackage[font=small]{caption}

\numberwithin{equation}{section}

\usepackage{verbatim}

\newcommand{\be}{\begin{equation}}
\newcommand{\ee}{\end{equation}}

\newcommand{\bml}{\begin{multline}}
\newcommand{\emll}{\end{multline}}

\newcommand{\nn}{\nonumber}

\def\({\left(} \def\){\right)}
\def\[{\left[} \def\]{\right]}

\def\mV{\mathcal{V}}

\def\al{\alpha}

\def\mA{\mathcal{A}}

\def\mM{\mathcal{M}}

\def\lam{\lambda}

\def\mW{\mathcal{W}}
\def\cW{\mathcal{W}}
\def\cV{\mathcal{V}}

\def\d{\partial}

\def\th{\theta}

\def\o{\omega}

\newcommand{\la}{\langle}
\newcommand{\ra}{\rangle}

\newcommand{\bi}{\begin{itemize}}
\newcommand{\ei}{\end{itemize}}

\newcommand{\bea}{\begin{eqnarray}}
\newcommand{\eea}{\end{eqnarray}}

\def\mU{\mathcal{U}}

\usepackage[left=2cm,right=2cm,top=2cm,bottom=2cm]{geometry}
\titleformat{\section}{\large\bfseries}{\thesection.}{4pt}{}
\titlespacing{\section}{0pt}{22pt}{6pt}

\titleformat{\subsection}{\large\bfseries}{\thesubsection.}{4pt}{}
\titlespacing{\subsection}{0pt}{18pt}{6pt}

\titleformat{\subsubsection}{\normalfont\bfseries}{\thesubsubsection.}{4pt}{}
\titlespacing{\subsubsection}{0pt}{16pt}{6pt}

\def\ie{\begin{equation}\begin{aligned}}
\def\fe{\end{aligned}\end{equation}}

\def\hat{\widehat}

\def\d{\partial}

\def\1{{\mathds 1}}

\def\mN{\mathcal{N}}
\def\mA{\mathcal{A}}

\def\o{\omega}

\DeclareFontShape{OT1}{cmr}{mx}{n}%
    {<->cmr10}{}
\newcommand{\mytitlefont}{\fontseries{mx}\selectfont}
\DeclareMathAlphabet{\titlemath}{OT1}{cmr}{mx}{n}

\def\mbS{\mathbb{S}}

\def\ss{\subsection}
\def\sss{\subsubsection}

\newcommand\zss[2]{\zeta_{\scaleto{#1}{5pt}}^{\scaleto{(#2)}{6pt}}}
\newcommand\zs[1]{\zeta^{\scaleto{(#1)}{6pt}}}
\newcommand\lamss[2]{\lambda_{\scaleto{#1}{3pt}}^{\scaleto{(#2)}{6pt}}}
\newcommand\lams[1]{\lambda^{\scaleto{(#1)}{6pt}}}

\newcommand\lamssp[2]{{\lambda'}_{\scaleto{#1}{3pt}}^{\scaleto{(#2)}{6pt}}}

\usepackage{scalerel}

\begin{document}

\begin{titlepage}

\begin{center}

~\\[1cm]

{\fontsize{20pt}{0pt} \mytitlefont Photon emission from an excited string}\\[10pt]

~\\[0.2cm]

{\fontsize{14pt}{0pt}Maurizio Firrotta {\small $^{1}$} and Vladimir Rosenhaus{\small $^{2}$}}

~\\[0.1cm]

\it{$^1$Department of Physics and INFN}\\ \it{ Tor Vergata University of Rome}\\ \it{ Via della Ricerca Scientifica 1, 00133, Rome, Italy}
\\[10pt]

\it{$^2$ Initiative for the Theoretical Sciences}\\ \it{ The Graduate Center, CUNY}\\ \it{
 365 Fifth Ave, New York, NY}\\[.5cm]

~\\[0.6cm]

\end{center}

\noindent 

We compute the amplitude for an excited string in any precisely specified state to decay into another excited string in any precisely specified state, via emission of a tachyon or photon. For generic and highly excited string states, the amplitude is a complicated function of the outgoing kinematic angle, sensitive to the precise state. We compute the square of this amplitude, averaged over  polarizations of the ingoing string and summed over  polarizations of the outgoing string. The seeming intractability of these calculations is made possible by extracting amplitudes involving excited strings from amplitudes involving  tachyons and a large number of  photons; the number of photons grows with the  complexity of the excited string state. Our work is in the spirit of the broad range of recent studies of statistical mechanics and chaos for quantum many-body systems.
The number of different excited string states at a given mass is exponentially large, and our calculation gives the  emission amplitude of a single photon from each of the microstates -- which, through the Horowitz-Polchinski correspondence principle, are in correspondence with  black hole microstates. 

\vfill

\end{titlepage}

\tableofcontents
~\\[.1cm]


\section{Introduction}

\begin{figure}[t]
\centering
\includegraphics[scale=0.5]{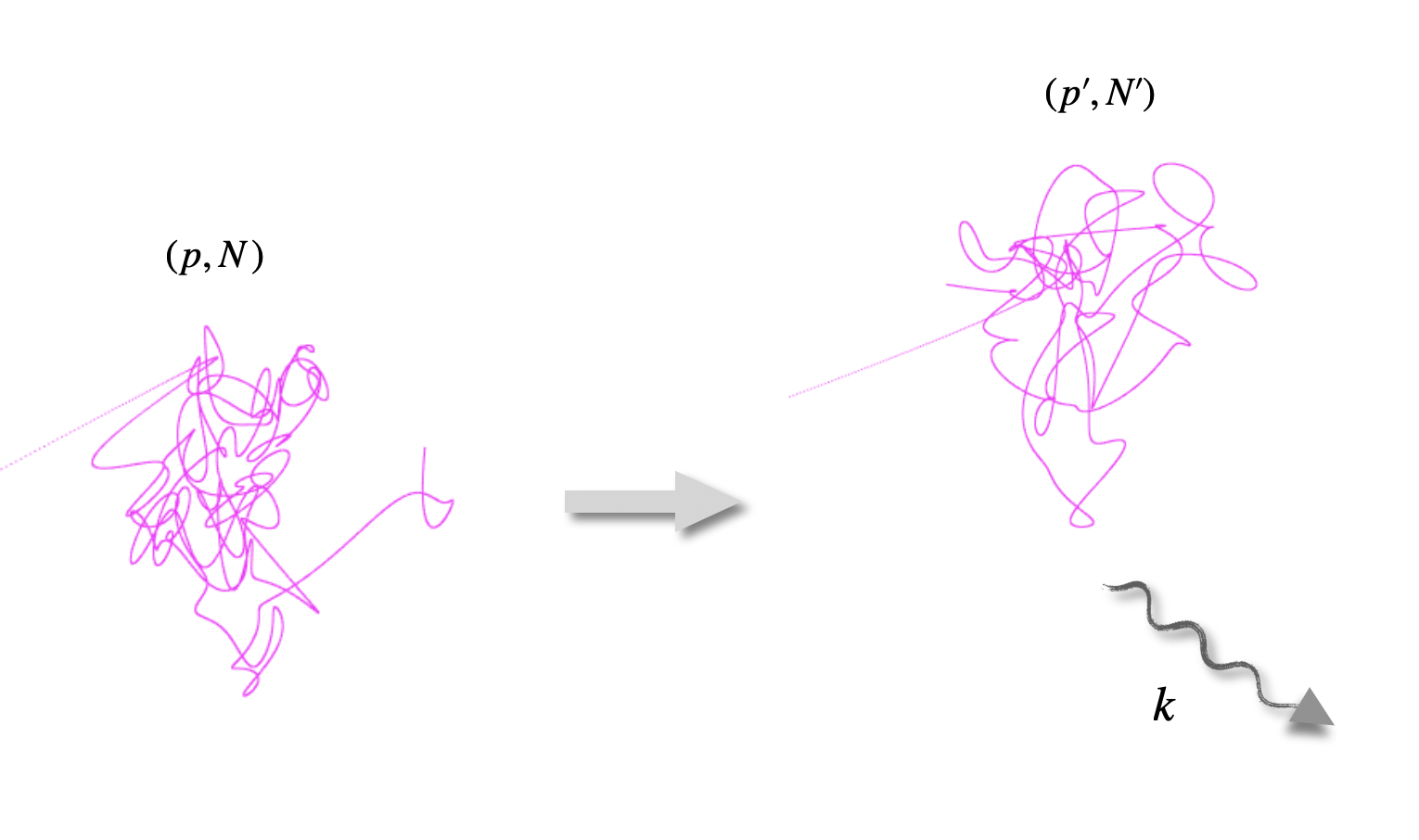}
\caption{The decay an excited string at level $N$ and momentum $p$ into an excited string at level $N'$ and momentum $p'$ via  emission of a tachyon (or a photon) of momentum $k$. We compute this amplitude, for any ingoing and outgoing excited string state. } \label{string_decay}
\end{figure}

This paper is about computing tree-level scattering amplitudes of excited strings in string theory \cite{Polchinski, GSW}. Concretely, we compute the amplitude for an excited string in any precisely specified state to decay into another excited string in any precisely specified state, via emission of a light string (a tachyon or a photon), see Fig.~\ref{string_decay}.  

The number of different states of a highly excited string grows exponentially with its mass, and the decay amplitude is a complicated function of the outgoing  angle,   depending sensitively on the precise state of the ingoing and outgoing string.

Our study is in part motivated by the large body of work, particularly in the last decade, broadly concerned with the emergence of thermality in quantum many-body systems and the  structure and characteristics of the underlying  microstates. This includes the eigenstate thermalization hypothesis \cite{Srednicki:1994mfb,Srednicki_1999,Rigol_2008,DAlessio:2015qtq}, its  intersection with  \cite{Murthy:2019fgs,Foini:2018sdb, Brenes:2021bjr, Pappalardi:2022aaz}  quantum many-body chaos and out-of-time-order correlators \cite{Kitaev, MSS, Blake:2022uyo}, and its realization in quantum field theory and high energy contexts, such as in \cite{Besken:2019bsu,Datta:2019jeo, Collier:2019weq, Delacretaz:2021ufg, Anous:2021caj,Karlsson:2021duj}.

Our specific thermodynamic system of interest is a highly excited string, and our goal is to extract the structure of the microstates through scattering amplitudes. Of course, a highly excited string is no ordinary quantum many-body system; plausibly, its microstates may account for a sizable fraction of the entropy of a black hole  \cite{HorowitzPolchinski,Horowitz:1997jc, Susskind:1993ws}. The process we compute -- emission of a photon from a highly excited string -- is of course the precise analog of a black hole, or a piece of coal, emitting a photon of radiation. 

This work is an outgrowth and continuation of the study of chaos in scattering amplitudes \cite{VRchaos, GR5, short}. More generally, the study of excited string scattering may connect with the large literature on modern scattering amplitudes in quantum field theory and string theory \cite{SAGE}, such as scattering of higher spin particles \cite{ Arkani-Hamed:2019ymq, Guevara:2019fsj}.

\begin{figure}[t]
\centering
\includegraphics[width=3.5in]{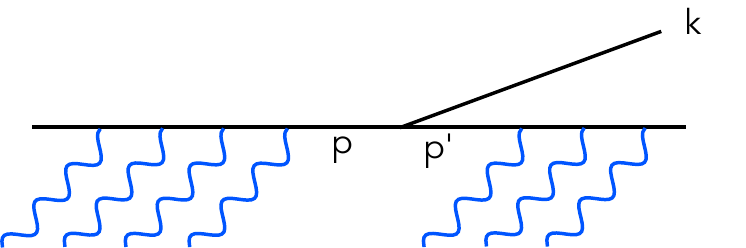}
\caption{  Scattering amplitudes involving excited strings, such as the one shown in Fig.~\ref{string_decay}, are most effectively computed from amplitudes involving exclusively tachyons and photons, with the excited strings appearing as intermediate states. } \label{IntroFig}
\end{figure}

Our calculation is made possible by  technical advances in string theory, fundamental yet recent, which dictate how to form highly excited strings from many photons. In short, any amplitude involving excited strings in any state can be extracted from an amplitude involving exclusively the light strings (tachyons and photons); the more complex the excited string states, the more photons need to be included, see Fig.~\ref{IntroFig}. For an amplitude with one excited string formed from photons of the same helicity this was shown explicitly in \cite{short}. 

More generally, important progress was made in \cite{Skliros}, which streamlined the approach of extracting excited string amplitudes from many-photon amplitudes, by finding explicit expressions for the appropriate excited string vertex operators \cite{DDF}. This result was recently used in \cite{GR5} to compute the amplitude involving one heavy string and two tachyons. Here we compute the more general amplitude, involving two heavy strings and one tachyon. More substantially,  to specify the state of the string one must not only specify which modes are excited (which partition of $N$ one is choosing), but also the polarizations (which, for a highly excited string, are complicated and constrained high rank tensors), for which there is no natural, generic choice. Of course, the standard approach in QED is to eliminate the photon polarization dependence by averaging the square of the amplitude over the polarizations. It is here that the power of viewing excited string amplitudes as coming from amplitudes of tachyons and photons really shines:  our problem of string polarization averaging reduces to the familiar and elementary task of photon polarization averaging. Implementing this gives us a simple expression.

\vspace{.5cm}

\noindent The paper is organized as follows: 

In Sec.~\ref{sec2} we  warmup with a  computation of the decay amplitude of the lightest massive string state into two tachyons. The computation is elementary, yet captures the essential physics which will appear in later sections for highly excited strings, without the notational complexity.

In Sec.~\ref{sec:31} we establish our notation for labelling excited strings. In Sec.~\ref{sec:kin} we briefly discuss how we find excited string amplitudes -- through amplitudes involving tachyons and photons, and we set up the kinematics for our decay process.  

In Sec.~\ref{sec4} we write down the amplitude and the polarization sum of the amplitude squared. In Sec.~\ref{sec41} we start with the case of an excited string decaying into two tachyons, while in Sec.~\ref{sec42} we study the more general case, of an excited string decaying into another excited string and emitting a tachyon.

 In Sec.~\ref{sec5} we take these results and look at  examples of the amplitudes for light strings.  
 
 We conclude in Sec.~\ref{sec6}.

Many of the technical details of the paper are relegated to the appendices. In Appendix~\ref{apa} we review the construction of vertex operators which create excited strings. These are used in Appendix~\ref{apb} to compute our amplitudes, and the sum over polarizations is performed in Appendix~\ref{apc}. In Appendix~\ref{apd} we indicate how the decay amplitudes studied in the main body of the text (which, despite the title of the paper, focuses on emission of a tachyon)  change if the excited string emits a photon instead of a tachyon.

\section{An invitation: lightest massive string decaying into two tachyons} \label{sec2}

We begin with a simple example, of the lightest massive string decaying into two tachyons. The lightest massive string is at level $N=2$, with mass $m^2=2$ and momentum $p$.~\footnote{We have set the square of the string length $\al'= 1/2$. All our calculations are for open strings in tree-level bosonic string theory in $D$ dimensions.   } It decays into a string at level $N'=0$ (a tachyon) of momentum $p'$ and mass $m^2=-2$, by emitting a tachyon with momentum $k$. The kinematics, with the massive string initially at rest, is, 
\bea \nn
p&=&\sqrt{2}(1, \vec 0)\\ \nn  
p'&=& -\frac{1}{\sqrt{2}}(1, \sqrt{5}\sin\theta, \sqrt{5}\cos\theta,  \vec 0) \\
k&=&- \frac{1}{\sqrt{2}}(1, -\sqrt{5}\sin\theta, -\sqrt{5}\cos\theta, \vec  0)~,
\eea
where $m^2 = -p^2 = 2$ and $p'^2=k^2=2$. 

The massive string can be in two different states: the state $(2)$ in which the second mode is excited $\lam{\cdot}A_{-2}|0\ra$, or the state $(1,1)$ in which the first mode is excited twice, $\lams{1}{\cdot}A_{-1} \lams{2}{\cdot}A_{-1}|0\ra$. These correspond to the two possible partitions of $N=2$. For each of these states there is a choice of polarizations, the $\lam$.~\footnote{Of course, this way of talking about states implies we are thinking in terms of light-cone gauge, in which $\lam{\cdot}A_{-2}|0\ra$ and $\lams{1}{\cdot}A_{-1} \lams{2}{\cdot}A_{-1}|0\ra$ are independent. In covariant gauge, these are part of one polarization tensor for the states at level $2$, see Appendix~\ref{apa}. In any case, for a string at a given level, we are free to compute amplitudes in which we average over  all or some or only one of the different states. The choice we will be making  is natural for light-cone gauge.}

\sss*{Ingoing string with second mode excited once}
\begin{figure}
\centering
\subfloat[]{\includegraphics[width=2.5in]{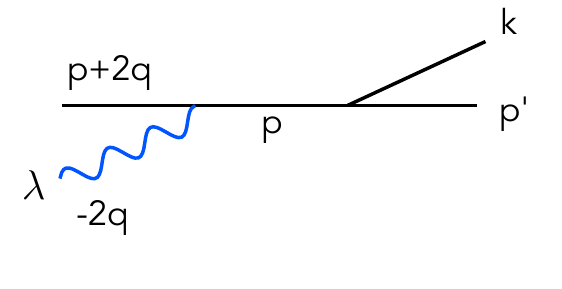}} \ \ \  \ \  \ \  \ \ \ \  \ \ \ 
\subfloat[]{\includegraphics[width=2.8in]{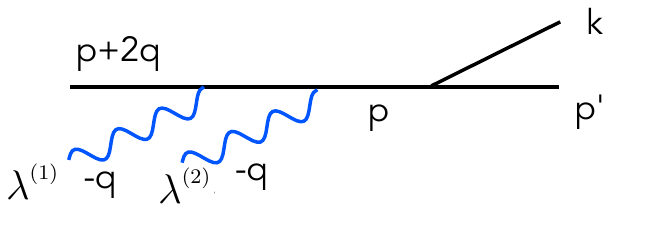}}
\caption{ (a) We extract the scattering amplitude involving an excited string in the state $\lam{\cdot}A_{-2}|0\ra$ from an amplitude involving a ``DDF photon'' of momentum $-2q$. (b) We extract the scattering amplitude involving an excited string in the state $\lams{1}{\cdot}A_{-1}\, \lams{2}{\cdot}A_{-1}|0\ra$ from an amplitude involving two ``DDF photons'', each of momentum $-q$. } \label{Ne2f}
\end{figure}

We start with the string in the state $(2)$. 

We would like to view amplitudes involving massive strings as arising from  scattering processes involving only tachyons and photons. Concretely, we start with a tachyon of momentum $p+2q$ and scatter a photon of momentum $-2q$ off of it, so that we produce our massive string of momentum $p$, see Fig.~\ref{Ne2f}(a). Since the tachyon must have mass $m^2 = -2$, we have the requirement that  $2=(p+2q)^2 =p^2 + 4 p \cdot q$ (where we used that the photon has null momentum, $q^2$=0) which means  that $q$ must satisfy $p {\cdot }q = 1$.  An acceptable choice of $q$ is, 
\be
q = -\frac{1}{\sqrt{2}}(1,0,1, \vec 0)~.
\ee
We will refer to this photon, of momentum $-2q$, as a ``DDF photon''. The photon will have a polarization $\lam$.  The polarization must be orthogonal to the momentum, $\lam {\cdot} q =0$ and, as is familiar, the amplitude is invariant under $\lam \rightarrow \lam+q$. There are therefore $D{-}2$ independent polarization vectors in $D$ dimensions. We take these to be the right and left circular polarizations, 
\be \label{lampm}
\lam_{\pm} = \frac{1}{\sqrt{2}}(0,1,0,\pm i,\vec 0 )~,
\ee
which are the complete set (and natural choice) in $D=4$, supplemented with $D-4$ other basis polarizations,  $\lam^I = (0, \ldots , 0, 1,0, \ldots, 0)$, where the $I$'th component is $1$ while all others are zero, if we are in $D>4$.~\footnote{Since the dynamics is confined to a three dimensional plane, giving special status to one of the transverse directions, as we have done in (\ref{lampm}), is artificial. We did this because our preference is to work in $D=4$, but for some questions it is important to work in the critical dimension of $D=26$.}
The amplitude we are interested in, of $p\rightarrow p' + k$, can be thought of as the residue of the pole of the amplitude for $(p+2q) + (- 2 q)\rightarrow p' + k$ at the resonance where the intermediate string is on-shell. 
It is useful to introduce the polarization $\zeta^{\mu}$, 
\be\label{26}
\zeta^{\mu} = \lam^{\mu} - (\lam \cdot p) q^{\mu}~.
\ee
This can be thought of as the polarization of our massive string: it manifestly satisfies $\zeta {\cdot} p =0$, because $q\cdot p =1$. For our kinematics, we see that $\zeta^{\mu} = \lam^{\mu}$. 
The amplitude for the decay $p\rightarrow p' + k$, denoted by ${\cal A}_{(2)(0)}$ to indicate that the initial string is in the state $(2)$ and the final string is in the state $(0)$ (a tachyon), will later be shown to be, 
\be \label{27}
{\cal A}_{(2)(0)}=-\frac{1}{\sqrt{2}}\zeta{\cdot}p'(1+2q{\cdot}p')~.
\ee
Taking the polarization to be  either left or right circular, as in (\ref{lampm}), and using the explicit kinematics we get the amplitude as a function of the angle $\theta$, 
\be
{\cal A}_{(2)(0)}^{(+)}={\cal A}_{(2)(0)}^{(-)}=-\frac{5}{2\sqrt{2}}\sin\theta\cos\theta~.
\ee
For the transverse polarizations $\lam^I$, the amplitude vanishes. 
In computing the scattering cross-section, it is common to average the  amplitude squared over polarizations. This is just one half of the sum over polarizations, 
\be \label{29}
\sum_{\text{polarizations}} |{\cal A}_{(2)(0)}|^2 = |{\cal A}_{(2)(0)}^{(+)}|^{2}+|{\cal A}_{(2)(0)}^{(-)}|^{2}={25\over 4}\sin^{2}\theta\cos^{2}\theta~,
\ee
where, as is familiar, we have done the sum over all polarizations by doing the sum over the  basis polarizations.

\sss*{Ingoing string with first mode excited twice}

Now let us look at the case in which the massive string is in the state $(1,1)$, in which the first mode is excited twice. To achieve this state, we take the scattering process in which we start with a tachyon of momentum $p+2q$, scatter a DDF photon with polarization $\lam^{(1)}$ and momentum $-q$ off of it, then scatter another DDF photon with polarization $\lam^{(2)}$ and momentum $-q$ off of it, and then get a decay into the tachyons with momenta $p'$ and $k$, see Fig.~\ref{Ne2f}(b). Looking at the residue of the pole where the internal strings are on-shell gives us the desired amplitude, 
\be \label{213}
{\cal A}_{(1,1),(0)}=\frac{1}{\sqrt{1+|{\zs{1}}^*\!{\cdot} \zs{2}|^2 }}\( \zs{1}{\cdot}p'\,\zs{2}{\cdot}p'+ {1\over 2}\zs{1}{\cdot}\zs{2}q{\cdot}p'(1+q{\cdot}p')\)~,
\ee
where, as in (\ref{26}), we introduced the polarizations $\zeta^{(1)}$ and $\zeta^{(2)}$,
\be
\zs{1}_{\mu} = \lams{1}_{\mu} - (\lams{1}{ \cdot} p) q_{\mu}~, \ \ \ \ \ \ \ \zs{2}_{\mu} = \lams{2}_{\mu} - (\lams{2}{ \cdot} p) q_{\mu}~.
\ee
Let us look at the amplitude for the different basis choices of polarization for the DDF photons. Each of the two photons can be either left or right circularly polarized (\ref{lampm}). The amplitudes are,
\be
{\cal A}_{(1,1)(0)}^{(+,+)}={\cal A}_{(1,1)(0)}^{(-,-)}=\frac{5}{4\sqrt{2}}\sin^{2}\theta\,, \quad 
{\cal A}_{(1,1)(0)}^{(+,-)}=\frac{5}{8}\sin^{2}\theta +\frac{1}{2}~, \ \ \ \ \ {\cal A}_{(1,1)(0)}^{(I,I)} = \frac{1}{8\sqrt{2}}(1 - 5 \cos^2 \theta)~,
\ee
in which the first/second superscript in the amplitude denotes the polarization of $\lams{1}$/ $\lams{2}$. 
Now squaring the amplitude and summing over the polarizations we have,
\be \label{sumpol20}
\sum_{\text{polarizations}} |{\cal A}_{(1,1)(0)}|^2  =2\(\frac{5}{4\sqrt{2}}\sin^{2}\theta\)^2 + \(\frac{5}{8}\sin^{2}\theta +\frac{1}{2}\)^2 + (D-4)\( \frac{1}{8\sqrt{2}}(1 - 5 \cos^2 \theta)\)^2~.
\ee

\sss*{Summary}
Let us summarize: we started with a massive string, $m^2=2$ (corresponding to level $N=2$), which decays into two tachyons. The amplitude depends on the state of the string, which  is characterized by which modes are excited and their polarizations. For this massive string at level two, one can either excite the second mode once (which we labelled as  $(2)$) or the first mode twice (which we labeled as $(1,1)$). For each of these choices, we computed the amplitude for a basis of polarizations and we computed the square of the amplitude and averaged over polarizations. 

If we wish, we may do a further sum over all the states at level $N=2$.  Setting the dimension to be the critical dimension, $D=26$, and summing (\ref{29})  and (\ref{sumpol20}) we get, 
\be
\sum_{\text{polarizations}} |{\cal A}_{(2)(0)}|^2+ \sum_{\text{polarizations}} |{\cal A}_{(1,1)(0)}|^2 = 3~.
\ee
The answer is independent of the angle -- as it should be, since a string at rest has spherical symmetry, if we sum over all possible polarizations.~\footnote{If we had not set $D=26$, there would be $\theta$ dependence. This is at it should be --  covariant gauge and light-cone gauge only agree for $D=26$, i.e. the string states, as counted in light-cone gauge (which is effectively what we are doing) only fall into representations of $SO(D{-}1)$ for $D=26$.}
\\[-2pt]

 The rest of the paper will be conceptually similar to this section. We will compute scattering amplitudes for the decay of an excited string (in the most general state) into another excited string (also in the most general state) through tachyon emission. We do this by forming the excited strings through repeated scatterings with photons. The number of photons we use and their polarizations determine the excited string state that we get. Viewing amplitudes involving 
excited strings as residues of poles of amplitudes involving only tachyons and photons is extremely useful: it provides a coherent wave of organizing the excited string states, it gives an effective computational tool for computing  amplitudes involving excited strings, and it allows the average over string polarizations to be almost trivially performed -- by  using the standard quantum field theory techniques for averaging over photon polarizations.

\section{Tachyon emission from a heavy string: setup} \label{sec3}

In Sec.~\ref{sec:31} we establish our notation for labelling excited strings. In Sec.~\ref{sec:kin} we briefly discuss how we find excited string amplitudes , through amplitudes with tachyons and photons. We  also set up the kinematics for our decay process,  which we will use in Sec.~\ref{sec5} when writing the amplitude as an explicit function of the angle $\theta$ of the outgoing excited string and tachyon.

\ss{Excited strings} \label{sec:31}
We have a  massive string with momentum $p$ and mass $-p^2 = 2(N{-}1)$ which decays into another massive string with mass $-p'^2 = 2(N'{-}1)$ and a tachyon of momentum $k$, $-k^2 =-2$, see Fig.~\ref{Kin}(a). 
The ingoing string is in the  excited state, 
\be \label{31}
\prod_{n,a} \lamss{n}{a}{\cdot}A_{-n} |0\ra~,
\ee
where $A_{-n}^{\mu}$ excites the $n$'th mode of the string in the $\mu$ spacetime direction and $\lam_n^{(a)}$ is the polarization. We let there be $g_n$ excitations of mode $n$; the product over $a$ above runs from $1$ to $g_n$ and the product over $n$ runs from one to infinity. If mode  a $n$ isn't excited, then its $g_n$ is zero, and that $n$ is excluded from the product. The string is at level $N$, which sets its mass,
\be
N=\sum_{n=1}^{\infty} n g_{n}~, \ \ \ \ \ -p^2 = 2(N-1)~.
\ee
There are a total of $J$ creation operators, $J = \sum_{n=1}^{\infty} g_n$. 

The excited string emits a tachyon and decays into an excited string in the state, 
\be \label{33}
\prod_{n,a} \lamssp{n}{a}{\cdot}A_{-n} |0\ra~,
\ee
where now the occupation number of level $n$ is $g_n'$, the polarizations are $ \lam'^{(a)}_n$, and the total level and mass are,
\be
 N'=\sum_{n=1}^{\infty} n g'_{n}~, \ \ \ \ \ \ -p'^2 = 2(N'-1)~.
\ee
There are a total of $J'$ creation operators, $J' = \sum_{n=1}^{\infty} g'_n$. 

If one is agnostic as to the polarizations, the state of a string at level $N$ is characterized by the partition of $N$. As is well known, the number of states at level $N$ -- which is the number of different partitions of $N$ -- grows exponentially with $\sqrt{N}$ for large $N$. We will label the state by the modes that are excited. For instance, at level $N=2$, we can either excite the first mode ($n=1$) twice, which we label as $(1,1)$, or we can excite the second mode ($n=2$) once, which we label as $(2)$. Likewise, for $N=3$ we have the states: $(3), (2,1), (1,1,1)$.

\subsection{Kinematics} \label{sec:kin}
\begin{figure}
\centering
\subfloat[]{\includegraphics[width=2.2in]{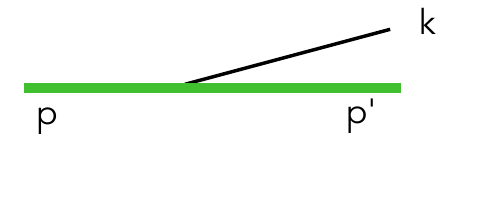}} \ \ \  \ \  \ \  \ \ \ \  \ \ \ 
\subfloat[]{\includegraphics[width=4in]{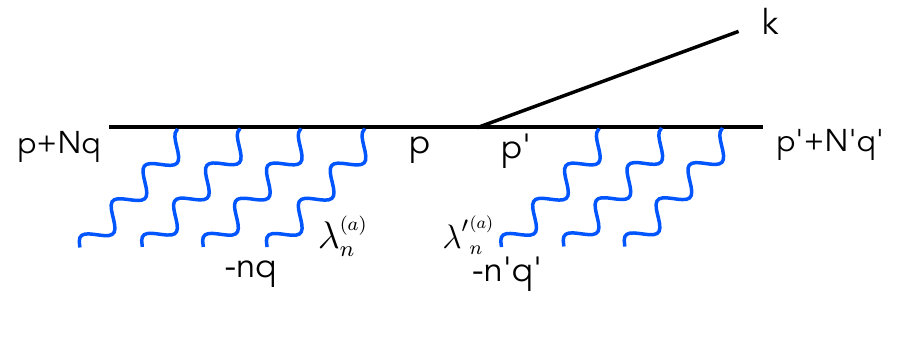}}
\caption{(a)We compute the amplitude of a string of momentum $p$,   in some definite excited state, to decay into another string of momentum $p'$ in a definite excited state, via emission of a tachyon of momentum $k$. The amplitude depends on the angle $\theta$ of the outgoing string and tachyon, and the polarizations of the ingoing and outgoing strings. (b) This amplitude is extracted from the amplitude involving a large number of ``DDF photons'' of momenta that are integer multiples of a null vector $q$. These photons can be thought of as creating the excited string. The precise state of the excited string sets the polarizations and momenta of the DDF photons.  } \label{Kin}
\end{figure}
For most of the discussion we will not need to pick an explicit frame of reference. However, to orient ourselves, it is useful to keep in mind the kinematics. 
It is convenient to choose kinematics in which the ingoing string is at rest. We take, 
 \bea \nn
 p&=&\sqrt{2N{-}2}(1,\vec{0})\\ \nn
 p'&=&-(E',\omega \sin\theta,\omega \cos\theta,\vec{0})\\
 k&=&-(E_{k},-\omega \sin\theta, -\omega \cos\theta,\vec{0})~, \label{35}
 \eea
where the energies are,
\be
 E'={N{+}N'{-}1\over \sqrt{2N{-}2}}~, \ \ \ \ \  \ \ \ \ \  \ \ \ \ E_{k}={N{-}N'{-}1\over \sqrt{2N{-}2}}~, \ \ \ \ \  \ \ \ \ \  \ \ \ \  \omega = \frac{\sqrt{(N{-}N'{-}1)^2 + 4(N{-}1)}}{\sqrt{2N{-}2}}~.
\ee
This satisfies momentum conservation, $p + p' + k = 0$. 

As mentioned earlier, we form an excited string by scattering so-called ``DDF photons'' off of a tachyon \cite{Skliros, DDF} (see also \cite{GR5, MF, MF2, MF3, MF4, MF6}). The DDF photons have momenta that are an integer multiple of $q$, where $q$ is a null vector,  $q^2=0$. Specifically, to create the ingoing excited string in the state (\ref{31}), we start with a tachyon with momentum $p+ N q$. We then successively scatter photons off of it --  one photon for each creation operator in (\ref{31}) -- that have momenta $-n q$ and polarization $\lam_n^{(a)}$. The same procedure applies to the outgoing excited string: it can be thought of as created from a tachyon of momentum $p'+N' q'$ off of which photons of momenta $-n q'$ and polarization $\lam'^{(a)}_n$ are scattered. See Fig.~\ref{Kin}(b). 

More precisely, we will look at the amplitude of $3{+}J{+}J'$ particles. Of these, three are tachyons: a tachyon of momentum $p{+}Nq$, a tachyon of momentum $p'{+}N' q'$, and a tachyon of momentum $k$. There are also $J$ photons: $g_n$ photons with momenta $-n q$ and polarizations  $\lam_n^{(a)}$ (with $a$ running from $1$ to $g_n$). And another $J'$  photons: $g_n'$ photons of momenta $-n q'$ and polarizations $\lam'^{(a)}_n$ (with $a$ running from $1$ to $g'_n$). The amplitude we are interested in -- of an excited string at level $N$ decaying into excited string at level $N'$ and a tachyon --  is the residue of the $J+J'$ order pole of this amplitude, where the intermediate strings are on-shell \cite{short}, 
\be \nn
\prod_{k=1}^J\prod_{k'=1}^{J'}\frac{\mA_{(m_1, m_2, \ldots, m_J)(m_1', m_2', \ldots, m_J')}}{\scriptstyle \[(p{+} N q {-}{\sum_{i=1}^k} m_i q)^2 {+} 2(\sum_{i=1}^k \!m_i {-} 1)\]\[(p'{+} N' q' {-} \sum_{i=1}^{k'} m_i' q')^2 {+} 2(\sum_{i=1}^{k'} \! m'_i {-} 1)\]} \sim \frac{\mA_{(m_1, m_2, \ldots, m_J)(m_1', m_2', \ldots, m_J')}}{(1-p{\cdot} q)^J (1- p'
{\cdot}q')^{J'}}~,
\ee
where the excited ingoing string is labelled as $(m_1, m_2, \ldots, m_J)$ and the excited outgoing string is labelled as $(m_1', m_2', \ldots, m_J')$. 

We see that (at the on-shell pole we are interested in) the DDF photons which create the ingoing massive string must have a $q$ for which $q{\cdot}p=1$ (this can equivalently be seen from the fact that we need $2 = (p+Nq)^2$). Likewise, the DDF photons forming the outgoing massive string have momenta which is an integer multiple of  $q'$ (which is  proportional to $q$)  for which $q'{\cdot}p'=1$.

Turning to the specific kinematics,  with our choice of kinematics above  the most general $q$ and $q'$ are, 
\be \label{39}
q={-1\over \sqrt{2N{-}2}}(1,\sin\beta,\cos\beta,\vec{0})~, \quad \quad q'={-1\over \omega \cos(\theta{-}\beta)-E'} (1,\sin\beta,\cos\beta,\vec{0})~,
\ee
where $\beta$ is an arbitrary angle. Since the amplitudes will only depend on $\theta{-}\beta$ we can, without loss of generality, set $\beta$ to be any value we like; we take $\beta=0$, 
\be \label{310}
q=-{1\over \sqrt{2N{-}2}}(1,0,1,\vec{0})~, \ \ \ \ \ \ \ \ \ \ \ \ \ \ \ \ \ \ q'=\frac{-1}{ \omega \cos\theta-E'} (1,0,1,\vec{0})~.
\ee

Let us look at the possible dot products of the various momenta: $p, p', k, q$. The dot products involving $p, p', k$ will all be constants. The  dot products involving $q$  are, 
\be \label{311}
q{\cdot } p' =\frac{1}{q'{\cdot p}} =-\frac{E'-\o \cos\th}{\sqrt{2N{-}2}}~.
\ee
This relation between $q{\cdot } p'$ and $q'{\cdot p}$ is general: using that $q'$ is proportional to $q$,  $q' = c\, q$ where $c$ is some constant, we can take the dot product with $p$  which gives $q'{\cdot} p = c$, or we can take the dot product with $p'$ which gives $1 = c\, q{\cdot} p'$. Finally, there is the dot product $k{\cdot} q$ which won't explicitly enter our amplitudes, but in any case follows from momentum conservation, $k{\cdot} q = -1 - p'{\cdot} q$. 

Each of the DDF photons will have a polarization. For the DDF photons forming the ingoing massive string the polarizations will be denoted by $\lam$, with potentially additional superscripts or subscripts to distinguish multiple DDF photons, and the DDF photons forming the outgoing massive string will have polarizations denoted by $\lam'$. The polarizations must be orthogonal to the momenta, $\lam{\cdot} q = \lam'{\cdot}q' = 0$, and thus take the form,
\be \label{312}
\lambda=\frac{1}{\sqrt{1+|\vec\Lambda|^{2}}}(0,1,0,\vec{\Lambda})~, \ \ \ \ \ \ \ \ \ \  \ \ \ \ \lambda'=\frac{1}{\sqrt{1+|\vec \Lambda'|^{2}}}(0,1,0,\vec{\Lambda'})~,
\ee 
with arbitrary $\vec \Lambda$ and $\vec \Lambda'$. The auxiliary polarization $\zeta$, see (\ref{26}), satisfies $\zeta {\cdot} p =0$ and is for our kinematics,
\be
\zeta^{\mu} = \lam^{\mu} - (\lam {\cdot} p) q^{\mu}  = \lam^{\mu}~,
\ee
because we have $\lam {\cdot} p=0$. Likewise, the auxiliary polarization for the outgoing DDF photons, $\zeta'^{\mu} =\lam'^{\mu} - (\lam' {\cdot} p') q'^{\mu}$, satisfies $\zeta'{\cdot} p=0$ and  for our kinematics is given by, 
\be
\zeta'^{\mu} = \frac{1}{\sqrt{1+|\vec \Lambda'|^{2}}}\(\frac{\o \sin\theta}{E'{-}\o \cos\theta},\, 1,\, \frac{-\o \sin\theta}{E'{-}\o \cos\theta}, \, \vec{\Lambda'}\)~.
\ee
We note that, 
\be
\zeta_n{\cdot} \zeta_m = \lam_n {\cdot} \lam_m~,
\ee
which is a coordinate invariant statement which follows from applying the definition (\ref{26}) of $\zeta$ and use of $\lam{\cdot} q = 0$. 

Finally, for the circular polarizations $\lam_{\pm} = \frac{1}{\sqrt{2}}(0,1,0, \pm i, \vec 0)$ and $\lam'_{\pm} = \frac{1}{\sqrt{2}}(0,1,0, \pm i, \vec 0)$ we have the following dot products for the corresponding $\zeta_{\pm}$ and $\zeta'_{\pm}$:  
\be
\zeta_{\pm} \cdot p' = -\frac{1}{\sqrt{2}} \o \sin \theta~, \ \ \ \zeta_{\pm}'{\cdot} p = \frac{ \o}{\sqrt{2}} q'{\cdot p} \sin \theta~.
\ee

Lastly, we  need a formula which expresses decay rates in terms of amplitudes. The familiar result in quantum field theory is, 
\be
d\Gamma = \frac{1}{2M}\( \prod_f \frac{d^{d-1}p_f}{(2\pi)^{d-1}} \frac{1}{2E_f}\) |\mA|^2 (2\pi)^d \delta^d (\sum p_i)~,
\ee
for the decay of a particle of mass $M$. For our kinematics, this becomes, 
\be
\frac{d\Gamma}{d\Omega}  = \frac{\omega^{d-3}}{16(N{-}1) (2\pi)^{d-2}}|\mA|^2~,
\ee
where $\Omega$ is the solid angle. For a given set of occupation numbers for the ingoing string and the outgoing string, we will want to compute the decay rate in which we are agnostic as to the polarizations: we want to average over the ingoing polarizations and sum over the outgoing polarizations. Averaging is proportional to summing, so the quantity we are interested in is  the sum of polarizations of the amplitude squared, 
\be
\sum_{\text{polarizations}}\!\!|\mA|^2~.
\ee
In what follows we will compute the amplitudes for states with definite occupation numbers and definite polarizations. In addition, for each set of occupation numbers, we will sum over all polarizations of the  square of the amplitude. 

\section{The amplitude} \label{sec4}
We now turn to computing the amplitude. 

\ss{Excited string decaying into two tachyons} \label{sec41}
Let us start with the special case in which the excited string decays into two tachyons ($N'=0$). This case is  slightly simpler than the case in which the outgoing string is excited. Also, for simplicity of notation, we mostly take the ingoing string to be in a state in which the first $k$ modes are each excited once -- the modifications to the formulas for  not exciting  some of these modes or exciting some of them multiple times is evident and will be given in Sec.~\ref{sec42}. 

\sss*{Amplitude}
We have an excited string in the state (see (\ref{31})) $\prod_n \lam_n {\cdot} A_{-n}|0\ra$ with momentum $p$, which was formed by scattering DDF photons of momenta $- n q$ and polarization $\lam_{n}$ off of a tachyon of momentum $p+Nq$. It decays into a tachyon of momentum $p'$ and a tachyon of momentum $k$. 
The amplitude is (see Appendix~\ref{apb} for the derivation), 
\be \label{41}
\mA = \zeta_{1,\mu_1} \cdots \zeta_{k, \mu_k} \mA^{\mu_1 \mu_2 \cdots \mu_k}~,
\ee
where 
$\zeta_{\mu_n}$ is defined in terms of the polarization $\lam_n$ for the $n$'th DDF photon, 
\be \label{42}
\zeta_{n, \mu} = \lam_{n, \mu} - (\lam_n {\cdot } p) q_{\mu}~,
\ee
and the tensor part of the amplitude is found by differentiating a generating function,
\be \label{316}
\mA^{\mu_1 \mu_2 \cdots \mu_k} = \frac{\d}{\d J_{1, \mu_1}}\cdots \frac{\d}{\d J_{k, \mu_k}}  \exp\Big(\sum_{n=1}^{\infty} J_{n}{\cdot}{\cal V}_{n}   + \sum_{1=n\leq m}^{\infty}  J_{n}{\cdot}J_{m} \,{\cal W}_{n,m} \Big)\Big|_{J_n=0}~.
\ee
The $\mV_{n}^{\mu}$ and $\mW_{n,m}$ that appear here are linear functions of  the outgoing tachyon momentum ($p'$) as well as nonlinear functions of its  scalar product with the DDF photon momentum,  $q{\cdot} p'$. Specifically, these functions are,
\be \label{VW}
\mV_{n}^{\mu}=-\frac{{p'}^{\mu} }{\sqrt{n}}\frac{(1+n q{\cdot}p')_{n-1}}{(n{-}1)!}~, \ \ \ \ \ \ \mW_{n,m}= \frac{n m}{n+m}\, q{\cdot }\mV_n\, q{\cdot} \mV_m \frac{q{\cdot}p' {+}1}{q{\cdot}p'}~,
\ee
where $(a)_n$ is the Pochhammer symbol, $(a)_n = a(a{+}1)\cdots (a{+}n{-}1)$ and 
which, explicitly, for some low values of $n$ are:
\be
{\cal V}_{1}^{\mu}= -p'^{\mu}~, \ \ \ \ {\cal V}_{2}^{\mu}=-\frac{p'^{\mu}}{\sqrt{2}}(1{+} 2 q{\cdot}p')~, \ \ \ \ \ {\cal V}_{3}^{\mu}=-\frac{p'^{\mu}}{\sqrt{3}}\frac{1}{2}(1{+} 3 q{\cdot}p')(2{+} 3 q{\cdot}p')~, \ \ \ \ \ \ \ \mW_{1,1} =  \frac{1}{2}q{\cdot}p'(1{+}  q{\cdot}p')~.
\ee
For a general ingoing string state, if some mode $n$ isn't excited, then we simply exclude the corresponding $\frac{\d}{\d J_{n, \mu_n}}$ term in (\ref{316}). Analogously, if a mode $n$ is excited multiple times then we add an additional superscript index to distinguish the different DDF photons at level $n$ (the polarizations are taken to be $\lam_n^{(a)}$), and we add in the corresponding number of additional derivatives in (\ref{316}), and modify the normalizations of the  $\lam_n^{(a)}$ so that the state has unit norm. In the event that this isn't clear, the next section will have an explicit formula with the additional notation. 

\sss*{Simple examples}
Let us look at some simple cases. 
In the simplest case, in which only mode $n$ is excited, the amplitude is, 
\be \label{319}
\mA =\zeta_{n} {\cdot} \mV_n~,
\ee
as is clear from taking a single derivative in (\ref{316}). 
Taking $n=2$ we recover our earlier result (\ref{27}). 
The next simplest case  is the one in which the ingoing string has only two modes, $n$ and $m$, which  are excited. The amplitude in this case  is,
\be \label{320}
\mA = \zeta_n {\cdot} \mV_n \, \zeta_m {\cdot} \mV_m+ \zeta_n{\cdot} \zeta_m \mW_{n,m}~,
\ee
which results from (\ref{316}), i.e. 
\be
\mA = \zeta_{n, \mu} \zeta_{m, \nu} \mA^{\mu \nu}~,\, \ \ \ \ \mA^{\mu \nu} = \frac{\d}{\d J_{n, \mu}} \frac{\d }{\d J_{m, \nu}} \exp\( J_n{\cdot}\mV_n {+} J_m{\cdot}\mV_m {+} J_n{\cdot} J_m \mW_{n,m}\)\Big|_{J_n=J_m = 0}~.
\ee
If $n=m$, then we add a superscript to distinguish the two polarizations, i.e., 
\be
\mA = \zeta_n^{\scaleto{(1)}{6pt}} {\cdot} \mV_n \, \zeta_n^{\scaleto{(2)}{6pt}} {\cdot} \mV_n+ \zeta_n^{\scaleto{(1)}{6pt}}{\cdot} \zeta_n^{\scaleto{(2)}{6pt}} \mW_{n,n}~.
\ee 
Setting $n=m=1$ gives the $\mA_{(1,1),(0)}$ which we wrote earlier in (\ref{213}). 
Analogously, if we excite three modes ($n, m, k$) the amplitude is, 
\be \label{322}
\mA = \zeta_n {\cdot }\cV_n \zeta_m {\cdot} \cV_m  \zeta_k {\cdot} \cV_k+\( \zeta_n{\cdot} \zeta_m \cW_{n,m} \zeta_k {\cdot} \cV_k+ \text{perm.}\)~.
\ee
The pattern is clear. 

A special case is when all the DDF photons are  left-circular polarized, or they are all  right-circular polarized. Then the dot products of all the polarizations are zero and the amplitude becomes, 
\be \label{411}
\mA^{(+, \ldots, +)}=\mA^{(-, \ldots, -)} = \prod_{n}\frac{1}{\sqrt{g_n!}}  (\zeta_+ {\cdot} \mV_{n})^{g_n}~.
\ee
Here we've added back in that mode $n$ can be excited $g_n$ times; i.e, we took the ingoing string to be in the state $\prod_n \frac{1}{\sqrt{g_n!}}(\lam_+ {\cdot} A_{-n})^{g_n} |0\ra$ where $\lam_{+}$ is the right-circular polarization (\ref{26}) and $\zeta_+$ is the corresponding $\zeta$, where recall that $\zeta_{\mu} = \lam_{\mu} - (\lam{\cdot} p)q_{\mu}$. In (\ref{411}) we are of course only taking the product over the $n$ for which $g_n$ is nonzero, so as to only include the state normalization factor $1/\sqrt{g_n!}$ for those modes. 

\sss*{Polarization average}
We have computed the amplitudes with an arbitrary polarization for the ingoing excited string. An obvious question is which polarization  we actually want. Indeed, we have a set of polarization vectors $\{\lam_n^{(a)}\}$, and each polarization vector can be an arbitrary superposition of the basis vectors. Specifying the polarizations is too much information. The usual approach in quantum field theory is to assume that we don't measure the polarization, and perform an average over all polarizations for the ingoing state. This is what we will do here. As we emphasized in the introduction, the advantage of our approach for computing amplitudes with excited strings -- by viewing them in terms of amplitudes having exclusively tachyons and photons -- is that it is straightforward to perform the sum over polarizations. 

Specifically, the scattering cross-sections involve the square of the amplitude. We  would like to  average  the square of the amplitude over all polarizations.  As is standard in quantum field theory, we do this by using that, within an amplitude,  $\sum_{\text{polarizations}} \lam_{\mu}^* \lam_{\nu} \rightarrow \eta_{\mu \nu}$. Doing this for our amplitude $\mA$ given in (\ref{41}--\ref{316}), we find that (see Appendix~\ref{apc}),
\be \label{323}
\sum_{\text{polarizations}}  |\mA|^2 = L_{\mu_1 \nu_1} L_{\mu_2 \nu_2} \cdots L_{\mu_k \nu_k}  \mA^{\mu_1 \mu_2 \cdots \mu_k} \mA^{\nu_1 \nu_2 \cdots \nu_k}~,
\ee
where we defined,
\be \label{54}
L_{\mu\nu}= \eta_{\mu \nu} - p_{\mu}q_{\nu} -p_{\nu}q_{\mu} + p^2 q_{\mu}q_{\nu}~.
\ee
$L_{\mu \nu}$  serves as a  projection operator and has the following properties:
 \be
 L_{\mu \nu}p^{\nu}  = L_{\mu \nu}q^{\nu}=0 ~, \ \ \ \ \ L_{\al \beta} L_{\rho \sigma} \eta^{\beta \sigma} = L_{\al \rho} \ \ \ \ \ L_{\mu\nu}=L_{\nu \mu}~, \ \ \ \ L_{\mu \nu} \eta^{\mu \nu}=L_{\mu \nu}L^{\mu \nu} =D{-}2~.
 \ee 
 The result (\ref{323}) is simple. We emphasize  that the averaging is over the $\lam$ polarizations which are the physical ones (and not the $\zeta$,  which are auxiliary vectors we have constructed). If in the amplitude we had had just $\lam$ instead of $\zeta$, then in (\ref{323}) we would have had $\eta_{\mu \nu}$ instead of $L_{\mu \nu}$. The $L_{\mu \nu}$ elegantly accounts for the $\zeta$.~\footnote{A simple way to see how $L_{\mu \nu}$ appears is the following. Recall that for a massive vector boson of momentum $p^{\mu}$ and mass $m$, inside the polarization sum of the square of the amplitude one has $\sum_{\lam}  \lam_{\mu}^* \lam_{\nu} \rightarrow P_{\mu \nu}$, where $P_{\mu \nu} = \eta_{\mu \nu}  - \frac{p_{\mu} p_{\nu}}{m^2}$ and has the property that $P_{\mu \nu}p^{\nu}=0$. This is a consequence of the orthogonality of the polarization  to the momentum, $\varepsilon_{\mu} p^{\mu} = 0$. Our particles have momenta that are $p$ plus different multiples of $q$. $L_{\mu \nu}$ achieves orthogonality regardless of what the multiple of $q$ is, by being orthogonal to both $p$ and $q$ individually.  }
 
 \vspace{.3cm}
 
 \noindent \textit{Simple Examples}
 
 Let us look at (\ref{323}) for some simple cases involving only a few excited modes, and see how it reproduces our earlier results. 
 For instance, if the ingoing string has only mode $n$ excited, the amplitude is (\ref{319}) and (\ref{323}) becomes,
 \be
 \sum_{\text{polarizations}}  |\mA|^2 =L_{\mu \nu} \mA^{\mu} \mA^{\mu}~, \ \ \ \ \ \mA^{\mu} = \mV_n^{\mu}~.
 \ee
 Taking $n=2$, we recover our earlier result (\ref{29}).
 
  Next, let us look at the case in which the ingoing string has  modes $n$ and $m$ each excited once. The amplitude is (\ref{320}) and (\ref{323}) becomes, 
 \be \label{417}
 \sum_{\text{polarizations}} |\mA|^2  =  L_{\mu_1 \nu_1} L_{\mu_2 \nu_2} \mA^{\mu_1 \mu_2} \mA^{\nu_1 \nu_2}~, \ \ \ \mA^{\mu \nu} = \mV_n^{\mu}\mV_m^{\nu} + \eta^{\mu \nu} \mW_{n,m}~.
 \ee
 Taking $n=m=1$ we recover the earlier result (\ref{sumpol20}). 
 
 As our last example, if we have an ingoing string in which modes $n$, $m$, and $k$  are each excited once, the amplitude is (\ref{322}) and  we see that (\ref{323}) gives, 
  \be
 \!\!\!   \sum_{\text{polarizations}}\!\!\!\!   |\mA|^2 = L_{\mu_1 \nu_1}  L_{\mu_2 \nu_2} L_{\mu_3 \nu_3} \mA^{\mu_1 \mu_2 \mu_3} \mA^{\nu_1 \nu_2 \nu_3}~, \ \ \ \ \mA^{\mu_1 \mu_2 \mu_3} = \mV_n^{\mu_1} \mV_m^{\mu_2} \mV_k^{\mu_3} + \( \mW_{n,m} \eta^{\mu_1 \mu_2} \mV_k^{\mu_3} + \text{perm.}\)
 \ee

To summarize this section: we have looked at the decay of an excited string at level $N$ with momentum $p$ into two tachyons of momenta $p'$ and $k$. The state of the excited string is parameterized by a choice of partition of $N$ (dictating which modes are excited) along with the polarizations $\lam_n^{(a)}$ of the excited modes, see (\ref{31}). The amplitude is given by (\ref{41}) and (\ref{316}), and the square of the amplitude summed over polarizations is given by (\ref{323}). 

\ss{Excited string decaying into an excited string and emitting a tachyon} \label{sec42}
We now turn to the more general case, of an excited string at level $N$ decaying into an excited string at level $N'$ through emission of a tachyon. As in the previous section (which took $N'=0$) the ingoing string is in the general state (\ref{31})  at level $N$, while the outgoing string is also in the general state (\ref{33}) at level $N'$. 

We are interested in the amplitude for arbitrary choices of polarization, and we are interested in the square of the amplitude summed over polarizations.

\sss*{Amplitude}
 We begin with the amplitude. We show in Appendix~\ref{apb} that the amplitude is, 
\be \label{419}
\mA =\prod_{n,a}\zeta^{\scaleto{(a)}{7pt}}_{n, \mu_{\scaleto{n}{3pt}}^{\scaleto{\! a}{3pt}} }\, {\zeta'}^{\scaleto{(a)}{7pt}}_{\!n, {\mu'}_{\scaleto{n}{3pt}}^{\scaleto{ a}{3pt}} }\, \,  \mA^{\mu_{\scaleto{1}{3pt}}^{\scaleto{\! 1}{3pt}} \cdots \mu'^{\scaleto{1}{3pt}}_{\scaleto{1}{3pt}}\cdots}~,
\ee
where, to be clear,  $\mA^{\mu_{\scaleto{1}{3pt}}^{\scaleto{\! 1}{3pt}} \cdots \mu'^{\scaleto{1}{3pt}}_{\scaleto{1}{3pt}}\cdots}$ contains $\sum_n g_n$ unprimed indices (one for each  polarization vector associated with the ingoing string) and $\sum_n g_n'$ primed indices (one for each  polarization vector associated with the outgoing string), and takes the form, 
\be \label{420}
\mA^{\mu_{\scaleto{1}{3pt}}^{\scaleto{\! 1}{3pt}} \cdots \mu'^{\scaleto{1}{3pt}}_{\scaleto{1}{3pt}}\cdots} = \prod_{n,a}\frac{\d }{\d J_{\scaleto{\! n}{3pt},\, \mu_{\scaleto{n}{2pt}}^{\scaleto{a}{2pt}}}^{\scaleto{(a)}{6pt}}}\frac{\d }{\d {J'}_{\!\scaleto{\! n}{3pt},\, {\mu\!'}_{\scaleto{n}{2pt}}^{\scaleto{a}{2pt}}}^{\scaleto{(a)}{6pt}}}\mA_{\text{gen}}\Big|_{J_{\scaleto{\! n}{3pt}, \mu_{\scaleto{n}{2pt}}^{\scaleto{a}{2pt}}}^{\scaleto{(a)}{6pt}}= {J'}_{\!\scaleto{\! n}{3pt}, {\mu\!'}_{\scaleto{n}{2pt}}^{\scaleto{a}{2pt}}}^{\scaleto{(a)}{6pt}}=  0}~,
 \ee
where $\mA_{\text{gen}}$ is,~\footnote{The explicit range of summation is:  in $\sum_{n,a}J_{n}^{(a)}{\cdot}{\cal V}_{n} $ we have  $1\leq n\leq \infty$ and $1\leq a\leq g_n$; in $\sum_{n,a}J'^{(a)}_{n}{\cdot}{\cal V}'_{n}$ we have $1\leq n\leq \infty$ and $1\leq a\leq g_n'$; in $\sum_{n,m,a,b} J_{n}^{(a)}{\cdot}J_{m}^{(b)} \,{\cal W}_{n,m}$ we have $1\leq n,m\leq \infty$ and $1\leq a\leq g_n$ and $1\leq b\leq g_m$, unless $a=b$, in which case we restrict to $n\leq m$, or $n=m$, in which case we restrict to $a\leq b$ (as we do not want  the same term to appear twice); a similar condition holds for $\sum_{n,m,a,b} J'^{(a)}_{n}{\cdot}J'^{(b)}_{m} \,{\cal W}'_{n,m} $ but with prime indices; and for $\sum_{n,m,a,b}J_{n}^{(a)}{\cdot}J'^{(b)}_{m}\,{\cal M}_{n,m} $ we have $1\leq n,m\leq \infty$ and $1\leq a\leq g_n$ and $1\leq b\leq g_m'$.} .  
 \be  \label{421}
\mA_{\text{gen}}\! =\! \exp\!\Bigg(\!\sum_{n,a} \(J_{n}^{(a)}{\cdot}{\cal V}_{n} +J'^{(a)}_{n}{\cdot}{\cal V}'_{n} \) + \sum_{n,m,a,b}\!\(\!  J_{n}^{(a)}{\cdot}J_{m}^{(b)} \,{\cal W}_{n,m} +J'^{(a)}_{n}{\cdot}J'^{(b)}_{m} \,{\cal W}'_{n,m} + J_{n}^{(a)}{\cdot}J'^{(b)}_{m}\,{\cal M}_{n,m} \)\!\! \Bigg)~.
\ee
 Here $\mV_n^{\mu}$ and $\mW_{n,m}$ were given earlier in (\ref{VW}). Exchanging $p$ with $p'$ and $q$ with $q'$, gives ${\cal V'}_{n}^{\mu}$ and $ \mW'_{n,m}$, 
\be \label{422}
{\cal V'}_{n}^{\mu}=(-1)^{n+1}\frac{{p}^{\mu}}{\sqrt{n}}\frac{(1+n q'{\cdot}p)_{n-1}}{(n{-}1)!}~, \ \ \ \ \ \ \mW'_{n,m}= \frac{n m}{n+m}\, q'{\cdot }{\cal V}'_n\, q'{\cdot} {\cal V}'_m \frac{q'{\cdot}p {+}1}{q'{\cdot}p}~.
\ee
Finally, the term arising from the contraction of the ingoing and the outgoing DDF photons is,
\be \label{423}
\mM_{n,m} =- q{\cdot }\mV_n\, q'{\cdot }{\cal V}'_m \frac{n m (1{+}q{\cdot}p' )}{m{+} n  q{\cdot}p' }~.
\ee
Notice that $\mM_{n,m}$ is invariant under $p, q, n  \leftrightarrow p', q', m$, as it should be. To see this, note that  $\frac{ 1{+}q{\cdot}p' }{m{+}  q{\cdot}p' } = \frac{ 1{+}q'{\cdot}p}{n{+}mq'{\cdot}p}$, where we used (\ref{311}): $q{\cdot}p' = 1/q'{\cdot} p$.

The amplitude is seemingly ambiguous,  because of the dependence on $q$ and $q'$, which are not uniquely specified. However as discussed earlier, below (\ref{39}), a rotation  of $q$ and $q'$ is simply a coordinate rotation, shifting the origin of $\theta$.

\sss*{Polarization   sum}
We now square the amplitude and sum over polarizations of the ingoing string and  the outgoing string. The result is similar to the one in the previous section, which had  $N'=0$ (\ref{323}), but with additional $L'_{ \mu \nu}$ factors coming from the polarizations of the outgoing string. In particular (see Appendix~\ref{apc}), 
the square of the amplitude summed over polarizations is, 
 \be  \label{424}
  \sum_{\text{polarizations}}|\mA|^2 = \prod_{n,a} L_{  \mu_{\scaleto{n}{2pt}}^{\scaleto{a}{2pt}}  \nu_{\scaleto{n}{2pt}}^{\scaleto{a}{2pt}} }\,   L'_{  {\mu'}_{\!\scaleto{n}{2pt}}^{\scaleto{a}{2pt}} { \nu'}_{\!\scaleto{n}{2pt}}^{\scaleto{a}{2pt}} } \,  \mA^{\mu_{\scaleto{1}{3pt}}^{\scaleto{\! 1}{3pt}} \cdots \mu'^{\scaleto{1}{3pt}}_{\scaleto{1}{3pt}}\cdots}  \mA^{\nu_{\scaleto{1}{3pt}}^{\scaleto{\! 1}{3pt}} \cdots \nu'^{\scaleto{1}{3pt}}_{\scaleto{1}{3pt}}\cdots} ~,
    \ee
    where $L_{\mu \nu}$ was given earlier in (\ref{54}) and 
$L'_{\mu \nu}$ is the projection operator for the outgoing string, defined analogously to $L_{\mu \nu}$, 
 \be
L'_{\mu\nu}= \eta_{\mu \nu} - p'_{\mu}q'_{\nu} -p'_{\nu}q'_{\mu} + p'^2 q'_{\mu}q'_{\nu}~.
\ee

\sss*{Simple examples}
To get oriented with the formula for the amplitude and the polarization-summed amplitude squared, let us look at some simple examples. 

First, we notice that if $N'=0$ we may set all the $J'^{(a)}_{n}$ terms in $\mA_{\text{gen}}$ to zero, and the amplitude reduces to what we had in Sec.~\ref{sec41} where we took $N'=0$ at the outset.
Notice also that the amplitude has a symmetry between excitations of the ingoing string and excitations of the outgoing string -- to get between ingoing and outgoing one adds a prime: $p,q, \zeta_{\mu} \leftrightarrow p',q', \zeta_{\mu}'$. 

Next, we look at  a simple example: consider exciting mode $n$ for the ingoing string and mode $m$ for the outgoing string. The amplitude (\ref{419}) is then, 
\be \label{426}
\mA =  \zeta_{n,\mu} \zeta_{m,\mu'}' \mA^{\mu \mu'}~, \ \ \  \sum_{\text{polarizations}} |\mA|^2  =  L_{\mu \nu} L'_{\mu' \nu'} \mA^{\mu \mu'} \mA^{\nu \nu'}~, \ \ \ \mA^{\mu \mu'} = \mV_n^{\mu}{\mV'}_m^{\mu'} + \eta^{\mu \mu'} \mM_{n,m}~.
\ee
As another example, if we excite modes $n$ and $m$ for the ingoing string and mode $k$ for the outgoing string, the amplitude is,
\be \nn
\mA =  \zeta_{n,\mu_1} \zeta_{m, \mu_2} \zeta_{k,\mu_1'}'  \mA^{\mu_1 \mu_2 \mu_1'} ~, \ \ \ \ \  \sum_{\text{polarizations}}   |\mA|^2 = L_{\mu_1\nu_1} L_{\mu_2\nu_2} L'_{\mu_1' \nu_1'} \mA^{\mu_1 \mu_2 \mu_1'}  \mA^{\nu_1 \nu_2 \nu_1'} 
\ee
\vspace{-.4cm}
\be \label{427}
  \mA^{\mu_1 \mu_2 \mu_1'} =\mV_n^{\mu_1} \mV_m^{\mu_2} {\mV'}_k^{\mu_1'} + \( \mW_{n,m} \eta^{\mu_1 \mu_2} {\mV'}_k^{\mu_1'} +\mM_{n, k} \eta^{\mu_1 \mu_1'} \mV_m^{\mu_2}+\mM_{m, k} \eta^{\mu_2 \mu_1'} \mV_n^{\mu_1} \)~.
 \ee
The pattern should be clear. 

Another special case is when all the DDF photons are  left-circular polarized (or they are all right-circular polarized). Then the dot products of all the polarizations are zero and the amplitude becomes, 
\be
\mA^{(+, \ldots, +) (+, \ldots, +)}=\mA^{(-, \ldots, -) (-, \ldots, -)} =   \prod_{n} \frac{1}{\sqrt{g_n!}} (\zeta_+ {\cdot} \mV_{n})^{g_n} \prod_n \frac{1}{\sqrt{g'_n!}}  (\zeta'_+{ \cdot} \mV'_{n})^{g_n'}~,
\ee
where  the first product is only over the $n$ for $g_n$ is nonzero, and the second product is only over the $n$ for which $g'_n$ is nonzero.

\section{Examples} \label{sec5}
In the previous section we found the amplitude $\mA$ (\ref{419}) for the decay of an excited string at level $N$ in an arbitrary state into another excited string at level $N'$ in an arbitrary state and a tachyon. We also found the  square of the amplitude, summed over polarizations of the ingoing string and polarizations of the outgoing string, (\ref{424}). Using the kinematics in Sec.~\ref{sec:kin} in which the ingoing string is at rest, these can be written as  functions of the angle $\theta$ of the outgoing string/tachyon. In this section we write the amplitude and the polarization summed  square amplitude   explicitly for various excited string states at low values of $N$ and $N'$.

The simplest case  is $N=1$ and $N'=0$ -- a photon decaying into two tachyons. We won't include it because it isn't compatible with our chosen kinematics of the initial string at rest. The next simplest case, $N=2$ and $N'=0$, was discussed in  Sec.~\ref{sec2}. So we begin with  $N=2$ and $N'=1$.

\ss{$N=2$, $N'=1$}
 We can partition $N=2$ in two ways: $(2)$ (the second string mode is excited once) and $(1,1)$ (the first string mode is excited twice). We look at each case in turn. 
 
\sss*{Ingoing state $(2)$}
The amplitude is given by (\ref{426}) with $n=2$ and $m=1$. Explicitly, 
\be
 {\cal A}_{(2)(1)}=  -\frac{1}{\sqrt{2}}(1+2q{\cdot}p')\zeta_{2}{\cdot}p' \zeta'_{1}{\cdot}p+ \sqrt{2}\zeta_{2}{\cdot}\zeta'_{1}(1+q{\cdot}p')~.
\ee
In the helicity basis the amplitude is,
\be
{\cal A}_{(2)(1)}^{(+)(+)}=-\frac{1}{\sqrt{2}}(\cos\theta + \cos (2\theta))~, \ \ \ \ \ {\cal A}_{(2)(1)}^{(+)(-)}=\frac{1}{\sqrt{2}}(\cos\theta -\cos (2\theta))~, \ \ \ \  {\cal A}_{(2)(1)}^{(I)(I)} =\sqrt{2} \cos \theta~. 
\ee
We have not written the other two choices of helicity, because the amplitudes are always invariant under a flip of all polarizations. So, ${\cal A}_{(2)(1)}^{(+)(+)}= {\cal A}_{(2)(1)}^{(-)(-)}$ and $ {\cal A}_{(2)(1)}^{(+)(-)} = {\cal A}_{(2)(1)}^{(-)(+)}$.
Squaring the amplitude and taking the sum over the polarizations (or directly using Eq.~\ref{424}) we get,
\be \label{53}
\sum_{\text{polarizations}}|{\cal A}_{(2)(1)}|^{2}= \cos 4\theta + (D{-}3)\cos 2\theta+ D{-}2~.
\ee

\sss*{Ingoing state $(1,1)$}
The amplitude is given by (\ref{427}) with $n=m=k=1$. Explicitly, 
\be
\mN^{-1}{\cal A}_{(1,1)(1)}=\zss{1}{1}{\cdot}p'\,\zss{1}{2}{\cdot}p'\,\zeta'_{1}{\cdot}p
-\zss{1}{2}{\cdot}p' \, \zss{1}{1}{\cdot}\zeta'_{1}-\zss{1}{1}{\cdot}p' \, \zss{1}{2}{\cdot}\zeta'_{1} + {1\over 2} \zss{1}{1}{\cdot}\zss{1}{2}\, \zeta'_{1}{\cdot}p \, q{\cdot}p' (1+ q{\cdot} p')~,
\ee
where the normalization factor is $\mN^{-2} =1+|{\zss{1}{1}}^*{\cdot} \zss{1}{2}|^2 $ (here and for what follows we will be assuming $|{\zss{1}{1}}|^2=|{\zss{1}{2}}|^2$=1). 
In the helicity basis the amplitude is, 
\be \nn
{\cal A}_{(1,1)(1)}^{(+,+)(+)}=\frac{-1}{\sqrt{2}}\sin \theta(1+\cos\theta)~, \ \ \ \  {\cal A}_{(1,1)(1)}^{(+,+)(-)}=\frac{1}{\sqrt{2}}\sin \theta(1-\cos\theta)~, \ \ \ \  
\ee
\vspace{-.8cm}
\be \nn
{\cal A}_{(1,1)(1)}^{(+,-)(+)}={\cal A}_{(1,1)(1)}^{(+,-)(-)}=-\frac{1}{4}\sin 2\theta~, \ \ \ {\cal A}_{(1,1)(1)}^{(+,I)(I)}=-\sin \theta~, \ \ \ \ {\cal A}_{(1,1)(1)}^{(I,I)(+)}=\frac{1}{4  \sqrt{2}} \sin2 \theta~.
   \ee
   The sum over polarizations is, 
\be \label{56}
\sum_{\text{polarizations}}|{\cal A}_{(1,1)(1)}|^{2}=\frac{1}{8}\sin^2 \theta\( (D{+}6)\cos 2\theta + 17D{-}42\)~.
\ee

As a check, notice that if we further sum over the ingoing states --  $(2)$ and $(1,1$)  --  we get,
\be
\sum_{\text{polarizations}}|{\cal A}_{(2)(1)}|^{2}+ \!\!\!\sum_{\text{polarizations}}|{\cal A}_{(1,1)(1)}|^{2}=\frac{1}{32}\( 65 D {-}154
 -(D{-}26) \cos 4\theta\)~,
 \ee
which is a constant for  $D=26$, as expected.

\ss{$N=3$, $N'=0$}
 We can partition $N=3$ in three ways: $(3)$ (the third string mode is excited once),  $(2,1)$ (the second string mode is excited once and the first string mode is excited once), $(1,1,1)$ (the first string modes is excited three times). We look at each case in turn. 

\sss*{Ingoing state $(3)$}
The amplitude is given by (\ref{319}) with $n=3$. Explicitly, 
\be
{\cal A}_{(3)(0)}={-1\over 2\sqrt{3}} (1+3q{\cdot}p')(2+3q{\cdot}p')\zeta_{3}{\cdot}p'~.
\ee
In the helicity basis the amplitude is, 
\be
{\cal A}_{(3)(0)}^{(+)}={-1 \over 8 \sqrt{2}}  \sin \theta \left(1-27 \cos ^2\theta \right)~.
\ee
   The sum over polarizations is, 
\be
\sum_{\text{polarizations}}|{\cal A}_{(3)(0)}|^{2}={1 \over 64} \sin^{2} \theta \left(1{-}27 \cos ^2\theta \right)^{2}~.
\ee

\sss*{Ingoing state $(2,1)$}
The amplitude is given by (\ref{320}) with $n=2$ and $m=1$. Explicitly, 
\be
{\cal A}_{(2,1)(0)}= (1+2q{\cdot}p')\zeta_{1}{\cdot}p'\,\zeta_{2}{\cdot}p'+{2\over 3}q{\cdot}p'(1+q{\cdot}p')(1+2q{\cdot}p')\zeta_{1}{\cdot}\zeta_{2}~.
\ee
In the helicity basis the amplitude is, 
\be \nn
{\cal A}_{(2,1)(0)}^{(++)}={3 \sqrt{3}\over 2\sqrt{2}}\sin^{2}\theta \cos\theta~, \ \ \ \ \ \ {\cal A}_{(2,1)(0)}^{(+-)}=\frac{7\cos\theta - 3\cos3\theta}{ 4\sqrt{6}} ~, \ \ \ \ \ \ {\cal A}_{(2,1)(0)}^{(II)} = \frac{\cos\theta(3 \cos 2\theta +1)}{4\sqrt{6}} ~.
\ee
   The sum over polarizations is, 
\be
\sum_{\text{polarizations}}|{\cal A}_{(2,1)(0)}|^{2}=  \frac{\cos^2\theta}{192} \( 9 (D{+}22) \cos 4\theta + 12(D{-}98)\cos 2\theta + 11D{+}914\)~.
\ee

\sss*{Ingoing state $(1,1,1)$}
The amplitude is given by (\ref{322}) with $n=m=k=1$ (and with the polarizations having a superscript distinguishing them). Explicitly, 
\be
\mN^{-1}{\cal A}_{(1,1,1)(0)}=-\zss{1}{1}{\cdot}p'\,\zss{1}{2}{\cdot}p'\,\zss{1}{3}{\cdot}p' - {1\over 2}q{\cdot}p'(1+q{\cdot}p')\left(\zss{1}{1}{\cdot}\zss{1}{2}\,\zss{1}{3}{\cdot}p' +\text{perm.} \right)~,
\ee 
where the normalization factor is,
\be
\mN^{-2} =1+|{\zss{1}{1}}^*{\cdot}\! \zss{1}{2}|^2 +|{\zss{1}{1}}^*\!{\cdot} \zss{1}{3}|^2 +|{\zss{1}{2}}^*\!{\cdot} \zss{1}{3}|^2 +( {\zss{1}{1}}^*\!{\cdot} \zss{1}{2} \, {\zss{1}{2}}^*\!{\cdot} \zss{1}{3}\,  {\zss{1}{3}}^*\!{\cdot} \zss{1}{1}+ \text{c.c.})~.
\ee
In the helicity basis the amplitude is, 
\be \nn
\!\! {\cal A}_{(1,1,1)(0)}^{(+++)}={3 \over4}\sin^{3}\theta~, \ \ \ \ \ {\cal A}_{(1,1,1)(0)}^{(++-)}=\frac{\sqrt{3} }{16} \sin\theta \left(7-3\cos2\theta \right)~, \ \ \ {\cal A}_{(1,1,1)(0)}^{(II+)}=\frac{\sqrt{3}}{64}(3 \sin 3\theta-\sin \theta)~.
\ee
   The sum over polarizations is, 
\be
\sum_{\text{polarizations}}|{\cal A}_{(1,1,1)(0)}|^{2}=\frac{3 \sin^2 \theta}{1024}\( (9D{+}48) \cos 4\theta + 12(D{-}48)\cos 2\theta + 11(D{+}48) \)~.
\ee

  As a check, notice that if we further sum over the ingoing states --  $(3)$, $(2,1)$ and $(1,1,1)$   --  we get, 
\bml
\sum_{\text{polarizations}}|{\cal A}_{(3)(0)}|^{2}+\!\!\!\!\! \sum_{\text{polarizations}}|{\cal A}_{(2,1)(0)}|^{2}+\!\!\!\!\! \sum_{\text{polarizations}}|{\cal A}_{(1,1,1)(0)}|^{2}\\=\frac{32668 + 634D}{12288} + \frac{D{-}26}{12288}\( 817 \cos 2\theta + 534 \cos 4\theta + 63 \cos 6\theta\)~,
\end{multline}
which is a constant for $D=26$, as expected.

\section{Discussion} \label{sec6}

Previous attempts to study highly excited strings have run up against the challenge of characterizing the high dimensional polarization tensor of the strings. This is overcome here, by building  excited strings out of photons. Our main result, given in Sec.~\ref{sec42}, is 
a well-organized expression for the amplitude  for  an excited string to decay into an excited string and a tachyon, as well as the decay rate for the process, which involves the square of the amplitude  averaged over ingoing polarizations and summed over outgoing polarizations.
 The expression is compact -- encoded in terms of derivatives of an exponential generating function, yet complex, reflecting the intricate structure of an excited string, whose state at mass level $N$ is characterized by the choice of partition of $N$.

We looked at explicit expressions for the amplitudes, as  functions of the kinematic angle, for various string states of low mass. The next step, which is our real interest, is to look in more detail at the  amplitudes for highly excited strings, and study the behavior of the microstates versus the ensemble. 
The microcanonical ensemble here is all highly excited strings at a given mass level. In fact, long ago  -- as a test of the correspondence principle between excited strings and black holes \cite{HorowitzPolchinski,Horowitz:1997jc, Susskind:1993ws} (see also \cite{StromingerVafa, Chen:2021b, Bena:2022rna})--  Amati and Russo \cite{Amati,   Manes3} computed the emission spectrum from an ensemble of excited strings, averaging over all strings at a given mass, finding the expected blackbody spectrum for a low energy photon emitted from the high mass string.   A fundamental set of questions throughout statistical mechanics is: how does the behavior of the microstates differ from the ensemble, how  is it that for many coarse-grained observables the microstate and the ensemble are indistinguishable, and which observables retain memory of the microstate? We are now equipped to address this in the context of a highly excited string -- an especially important quantum many-body system --  emitting a single tachyon or photon. We hope to report on this in future work. 

\sss*{Acknowledgments} This work was supported in part by NSF grant 2209116. 

\appendix

\section{Excited string vertex operators} \label{apa}

In this appendix we review the DDF vertex operators as constructed in \cite{Skliros}, see also \cite{GR5, MF}. 

Recall that the vertex operator for the tachyon is $e^{i p{\cdot} X}$, whereas the vertex operator for  an excited string takes the form of a polynomial of $\d^k X^{\mu}$ for various powers of $k$. Specifically, the DDF vertex operator is a polynomial made up of $q{\cdot} \d^k X$ and  $\zeta{ \cdot} \d^k X$ for various $k$. 

We first write down the DDF vertex operator which creates a  string in which mode $n$ is excited, $\frac{1}{\sqrt{n}}\lam_n{\cdot} A_{-n}|0\ra$. The expression will require some unpacking. The vertex operator is, 
\be
V_{(n)}(z) = \zeta_n {\cdot} P_{n}\, e^{i p \cdot X}~,
\ee
where, as in the main body of the text, $\zeta_{n, \mu} = \lam_{n, \mu} - (\lam_n{\cdot}p)q_{\mu}$ and, 
\be \label{A2}
 P_n^{\mu}(z) =\frac{1}{\sqrt{n}} \sum_{m=1}^n\frac{i}{(m{-}1)!} \d^m X^{\mu}\, S_{n-m}(\mU_r^{(n)})~.
\ee
Here $S_n(u_r)$ is a function of the set of variables $\{u_r\}$ with integer $r$, which is defined by a contour integral, 
\be
S_n(u_r)  =\oint_0\frac{d w}{2\pi i} \frac{1}{w^{n+1}} \exp\(\sum_{r=1}^n u_r w^r\)~.
\ee
To obtain the explicit expression for a given $n$, one can Taylor expand the exponential to pick out the $w^n$ term, which is the one that contributes. For instance, for the first three values of $n$,  
\be
S_1(u_r) = u_1~, \ \ \ S_2(u_r) = \frac{1}{2}u_1^2 + u_2~, \ \ \ S_3(u_r) = \frac{1}{6}u_1^3 + u_1 u_2 + u_3~.
\ee
For our expression (\ref{A2}) the variables $u_r$ are $\mU_r^{(n)}$  (the $r$ denotes the label of the set; the $n$ is an additional index), 
\be
\mU_r^{(n)} = -n\frac{i }{r!} q{\cdot}\d^r X~.
\ee
Combining everything, we see that the DDF vertex operator is, for instance, for $n=2$, 
\be \label{A6}
V_{(2)} =\frac{1}{\sqrt{2}} \( i \zeta_2 {\cdot }\d^2 X + 2 (\zeta_2 {\cdot} \d X) ( q {\cdot} \d X) \) e^{i p \cdot X}~.
\ee

Let us check that this is correct. The standard way of constructing vertex operators is in the covariant formalism. One looks for operators annihilated by the Virasoro generators (corresponding to physical states; reflecting diffeomorphism invariance of the string worldsheet). At level-two this gives the vertex operator,
\be \label{A7}
\(i \xi_{\mu} \d^2 X^{\mu} + \xi_{\mu \nu} \d X^{\mu} \d X^{\nu}\) e^{i p \cdot X}~,
\ee
where the Virasoro constraints require that the polarization vector $\xi_{\mu}$ and tensor $\xi_{\mu \nu}$ satisfy, 
\be \label{B9}
\xi_{\mu} - p^{\nu} \xi_{\mu \nu} = 0~, \ \ \ \ \eta^{\mu \nu} \xi_{\mu \nu} -2 p{\cdot} \xi = 0~.
\ee
Our DDF vertex operator (\ref{A6}) is precisely of this form, with $\xi_{\mu} = \frac{1}{\sqrt{2}}\zeta_{2, \mu}$ and 
$\xi_{\mu \nu} = \frac{1}{\sqrt{2}}(\zeta_{2, \mu } q_{\nu} + \zeta_{2,\nu} q_{\mu})$. The Virasoro constraints are automatically satisfied. This is one of the major advantages of using DDF vertex operators, which becomes particularly apparent at high level: as opposed to covariant vertex operators, in which the polarization tensors have a sequence of Virasoro constraints they must satisfy, for the DDF vertex operators the constraints are automatically satisfied.

Now let us look at the DDF vertex operator for two excited modes, corresponding to the state $\frac{1}{\sqrt{nm}}(\lam_n{\cdot} A_{-n})(\lam_m {\cdot}A_{-m})|0\ra$. It is given by, 
\be  \label{339}
V_{(n,m)}  = \( \zeta_n {\cdot} P_{n}\, \zeta_m {\cdot} P_{m} + \zeta_n {\cdot} \zeta_m \, \mbS_{n,m} \)   e^{i p \cdot X}~,
\ee
where the term associated with the dot product of the polarizations is defined as,
\be \label{A10}
\mbS_{n,m} =\frac{1}{\sqrt{nm}} \sum_{l=1}^m l\, S_{n-l}(\mU_r^{(n)})S_{m+l}(\mU_r^{(m)})~, \  \ \ \text{for} \ \ \ m\geq n~, \ \ \ \mbS_{m, n} = \mbS_{n,m}~.
\ee
For instance, for $n=m=1$, 
\be
V_{(1,1)}= - \Big( \zss{1}{1}{\cdot} \d X\,  \zss{1}{2} {\cdot} \d X + \frac{1}{2} \zss{1}{1}{ \cdot} \zss{1}{2}\( i q{\cdot }\d^2 X + (q{\cdot} \d X)^2\) \Big)e^{i p \cdot X}~.
\ee
This is consistent with the vertex operator in  covariant form (\ref{A7}), 
with the identification,
\be \label{324}
\xi_{\mu} =-\frac{1}{2} \zss{1}{1} {\cdot} \zss{1}{2}\, q_{\mu}~, \ \ \ \ \ \ \ \ \  \ \ \xi_{\mu \nu} =-\frac{1}{2}\( \zss{1}{1}{\cdot }\zss{1}{2}\,  q_{\mu} q_{\nu} +\zss{1, \mu}{1}\zss{1, \nu}{2}+ \zss{1, \nu}{1}\zss{1, \mu}{2}\)~.
\ee
Again, the Virasoro constraints (\ref{B9}) are automatically satisfied. 

Finally, the DDF vertex operator for the general state $\prod_{n=1}^{\infty}\prod_{ a=1}^{g_n}\frac{1}{\sqrt{n}} \lam_n^{(a)}{\cdot} A_{-n}|0\ra$ is,~\footnote{The explicit range of summation in the exponent of (\ref{vert}) is:  in the first term we have  $1\leq n\leq \infty$ and $1\leq a\leq g_n$; in the second term we have $1\leq n,m\leq \infty$ and $1\leq a\leq g_n$ and $1\leq b\leq g_m$, unless $a=b$, in which case we restrict to $n\leq m$, or $n=m$, in which case we restrict to $a\leq b$ (as we do not want the same term to appear twice).}
\be \label{vert}
V= e^{i p{\cdot} X} \prod_{n, a} \zeta_n^{(a)}{\cdot} \frac{\d}{\d J_n^{(a)}}\,  \exp\( \sum_{n,a} J_n^{(a)} {\cdot} P_n + \sum_{n,m, a,b}\!\! J_n^{(a)} {\cdot} J_m^{(b)}\, \mbS_{n, m} \)\Big|_{J=0}~.
\ee
In short, every polarization $\zeta_n^{(a)}$ must appear one, and it can  appear either by being contracted into $P_n$, or by being contracted with some other polarization $\zeta_m^{(b)}$ and come with a factor of $\mbS_{n, m}$. 

Let us briefly comment on the normalization of the states, and correspondingly of the vertex operators. The state with a single creation operator, $\frac{1}{\sqrt{n}} \lam_n {\cdot} A_{-n}|0\ra$, has the inner product $|\lam_n|^2$, as a result of the string theory conventions for the commutation relations, $\[ A_{n}, A_{-m}\] = n \delta_{n, m}$. This is the origin of the $1/\sqrt{n}$ factor in $P_n^{\mu}$ in (\ref{A2}) and $\mbS_{n, m}$ in (\ref{A10}).  Furthermore, we would like the state to have inner product equal to one,  so we pick the normalization of $\lam_n$ so that $|\lam_n|^2$=1. Notice that if we have a state in which we excite the same mode twice, such as,
\be
\frac{1}{n} (\lamss{n}{1} {\cdot}A_{-n})(\lamss{n}{2} {\cdot}A_{-n}) |0\ra~, 
\ee
then we pick the polarizations to have the normalization $|\lamss{n}{1}|^2 |\lamss{n}{2}|^2 + |{\lamss{n}{1}}^*\!{\cdot} \lamss{n}{2}|^2 = 1$.

\section{Derivation of the amplitude} \label{apb}

We will compute a three-point amplitude, with an excited string of momentum $p$, another excited string of momentum $p'$, and a tachyon of momentum $q$. The amplitude is given by, 
 \be \label{B1}
 \mA = \la V(z) V'(z')  e^{i k{ \cdot }X(w) }\ra~,
 \ee
 where the vertex operator for the excited string of momentum $p$ sits at $z$ and was given in Appendix~\ref{apa}, (\ref{vert}), the vertex operator for the excited string of momentum $p'$ sits at $z'$ and is given by (\ref{vert}) but with primed variables instead of unprimed variables, $p, \zeta, q \rightarrow p', \zeta', q'$, and the tachyon sits at $w$. By SL$_2$ invariance, we are allowed to choose any $z, z', w$, and after computing the correlator we will take,  
 \be \label{points}
z=0~, \ \ \ \ z' = 1~, \ \ \ \ \ w = \infty~.
\ee
 Using the explicit form of the vertex operator (\ref{vert}), we may write the amplitude as, 
 \be
 \mA = \prod_{n,a} \zeta_n^{(a)}{\cdot} \frac{\d}{\d {J}_n^{(a)}} \, {\zeta'}_n^{(a)}{\cdot} \frac{\d}{\d {J'}_n^{(a)}} \mA_{\text{gen}}\Big|_{J = J'=0}~,
 \ee
 where,
\bml
\mA_{\text{gen}}\! =\Big \langle \exp\!\Big(\!\sum_{n,a} \(J_{n}^{(a)}{\cdot}{P}_{n}(z) {+}J'^{(a)}_{n}{\cdot}P'_{n}(z') \){ +}\!\! \sum_{n,m,a,b}\!\!\(\!  J_{n}^{(a)}{\cdot}J_{m}^{(b)} \,{\mbS}_{n,m}(z) {+}J'^{(a)}_{n}{\cdot}J'^{(b)}_{m} \,{\mbS}'_{n,m}(z') \)\!\! \Big) \\
e^{i p{\cdot}X(z)} e^{ i p'{\cdot} X(z')}e^{i k{ \cdot }X(w)}\Big \rangle ~.
\end{multline}
We now need to compute this correlation function. This is straightforward, since $X^{\mu}$ is a free field with the  two-point function, 
\be \label{B5}
\la X^{\mu}(z_1)\d X^{\nu}(z_2)\ra = \frac{\eta^{\mu \nu}}{z_{12}}~.
\ee
We simply need to perform the Wick contractions. We start with  $P_n(z)$, defined in (\ref{A2}). We look at the terms in which it contracts with the exponentials $e^{ i p'{\cdot} X(z')}e^{i k{ \cdot }X(w)}$. We may  make the replacement, 
\be
\frac{i}{(m{-}1)!}\d^m X^{\mu} \rightarrow \frac{-{p'}^{\mu}}{(z'-z)^m} - \frac{k^{\mu}}{(w{-}z)^m} \rightarrow -{p'}^{\mu}~,
\ee
where for the second arrow, we set the points to be (\ref{points}). Likewise, we may make the replacement, $\mU_r^{(n)} (z) \rightarrow {\hat\mU}_r^{(n)}$ where,
\be
{\hat \mU}_r^{(n)} = \frac{n}{r} \(\frac{p'{\cdot}q }{(z'{-}z)^r} + \frac{k{\cdot}q }{(w{-}z)^r} \) \rightarrow \frac{n}{r} p'{\cdot} q~,
\ee
where for the second replacement we set the points to be (\ref{points}).
Finally, since
\be \label{SmP}
S_m\(  \frac{n}{r} p'{\cdot} q\) = \frac{(n p'{\cdot} q)_m}{m!}~,
\ee
we have that inside the expectation value,
\be
P_n^{\mu} (z) \rightarrow- \frac{{p'}^{\mu}}{\sqrt{n}} \sum_{m=1}^n S_{n-m}\(  \frac{n}{r} p'{\cdot} q\) =
  \mV_n^{\mu}~,
  \ee
  where $ \mV_n^{\mu}$ was defined in (\ref{VW}). 
  Turning to $\mbS_{n,m} $ defined in (\ref{A10}), using (\ref{SmP}) we have that inside the expectation value,
\be
\mbS_{n,m} \rightarrow\frac{1}{\sqrt{nm}} \sum_{l=1}^m l\, S_{n-l}\(\frac{n}{r} p'{\cdot} q\)S_{m+l}\(\frac{m}{r} p'{\cdot} q\) = \mW_{n,m}~,
\ee
where we did the sum over $l$, and $\mW_{n,m}$ was defined in (\ref{VW})

Similar results carry over for the primed variables. Namely, we may make the replacement,
\be
\frac{i}{(m{-}1)!}\d^m X^{\mu} \rightarrow \frac{-{p}^{\mu}}{(z-z')^m} - \frac{k^{\mu}}{(w{-}z')^m} \rightarrow (-1)^{m+1}{p}^{\mu}~,
\ee
where for the second arrow, we set the points to be (\ref{points}). Likewise, we make the replacement, 
 ${\mU'}_r^{(n)} (z') \rightarrow {\hat{\mU}_r^{(n)}}$ where,
 \be
 {\hat{\mU}_r^{(n)}}= \frac{n}{r} \(\frac{p{\cdot}q' }{(z{-}z')^r} + \frac{k{\cdot}q' }{(w{-}z')^r} \) \rightarrow (-1)^r\frac{n}{r} p{\cdot} q'~.
\ee
Now since, 
\be \label{SmP2}
S_m\( {\hat{\mU}_r^{(n)}}\) =(-1)^m \frac{(n p{\cdot} q')_m}{m!}~,
\ee
inside the expectation value we have, 
\be
{P'}_n^{\mu} (z') \rightarrow \frac{{p}^{\mu} }{\sqrt{n}}\sum_{m=1}^n(-1)^{m+1}(-1)^{n-m} S_{n-m}\(  \frac{n}{r} p{\cdot} q'\) =
   {\mV'}_n^{\mu}~,
  \ee
  with $ {\mV'}_n^{\mu}$ defined in (\ref{422}). Likewise, $\mbS'_{n,m}\rightarrow \mW'_{n,m}$, also defined in (\ref{422}).

Finally, we need to look at the term coming from the contraction of $P_n(z)$ and $P_m(z')$. Using (\ref{A2}) we get, 
\be
\la P_n^{\mu}(z) P_m^{\nu} (z') \ra =\eta^{\mu \nu}\frac{1}{\sqrt{nm}} \sum_{l_1, l_2} \frac{(-1)^{l_1+1} (l_1{+}l_2{-}1)!}{(l_1{-}1)!(l_2{-}1)!} \frac{1}{(z-z')^{l_1 + l_2}}S_{n-l_1}\( {\hat{\mU}_r^{(n)}}\) S_{m-l_2}\Big( {\hat{\mU'}}_r^{(n)}\Big)~,
\ee
where we used (\ref{B5}), which upon differentiating gives, 
\be
\la \d^{l_1} X^{\mu}(z)\d^{l_2}X^{\nu}(z') \ra = (-1)^{l_1} \frac{(l_1 {+} l_2{-}1)!}{(z-z')^{l_1{+}l_2}}~.
\ee
Taking the points to be at locations (\ref{points}) we get, 
\be
\la P_n^{\mu}(z) P_m^{\nu} (z') \ra = \eta^{\mu \nu}\frac{(-1)^{m+1}}{\sqrt{nm}}\sum_{l_1, l_2} \frac{ (l_1{+}l_2{-}1)!}{(l_1{-}1)!(l_2{-}1)!} \frac{(n p'{\cdot}q )_{n-l_1}}{(n{-}l_1)!} \frac{(m p{\cdot}q' )_{m-l_2}}{(m{-}l_2)!}  = \eta^{\mu \nu} \mM_{n, m}~,
\ee
where $ \mM_{n, m}$ was defined in (\ref{423}) and we used (\ref{SmP}) and (\ref{SmP2}) and that $q{\cdot} p' = 1/q'{\cdot} p$. 

In total, we have recovered the amplitude claimed in the main body of the text, (\ref{420}) and (\ref{421}). 
 
\subsection{Averaging over polarizations} \label{apc}
In this appendix we derive the sum over DDF photon polarizations of the square of the amplitude given in the main body of the text, (\ref{424}). 

We start by recalling the standard result in quantum field theory, that one may replace a sum over photon polarizations as $\sum_{\lam} \lam^*_{\mu} \lam_{\nu} \rightarrow \eta_{\mu \nu}$. This is easy to see. Let us write the amplitude as $\mA = \lam_{\mu} \mA^{\mu}$, where we have separated out the dependence of the amplitude on the polarization $\lam_{\mu}$ of some particular photon.
We take a basis of photon polarizations (see (\ref{312})) to be $ \lam = (0,1,0,0)$ and $\lam = (0,0,0,1)$, where we have restricted to four dimensions for simplicity. This gives for the sum, 
\be \label{lamS}
\sum_{\lam} |\mA|^2 = \sum_{\lam} \lam_{\mu}^* \lam_{\nu} \mA^{\mu} \mA^{\nu} = |\mA^1|^2 + |\mA^3|^2~.
\ee
However, the amplitude is invariant under $\lam_{\mu} \rightarrow \lam_{\mu} + q_{\mu}$ (the photon momentum is proportional to $q_{\mu}$), and so $q_{\mu}A^{\mu} = 0$. With our $q_{\mu}$ in (\ref{310}) this gives $\mA^0 +\mA^2$. Consequently, $-|\mA^{0}|^2 + |\mA^2|^2 = 0$ and we may replace (\ref{lamS}) with, 
\be
\sum_{\lam} |\mA|^2 = \eta_{\mu \nu} \mA^{\mu} \mA^{\nu}~,
\ee
as desired. 

Now we need to apply this result to our amplitude (\ref{420}) for the decay of an excited string into another excited string and a tachyon. 

To start, let us look at the amplitude for an ingoing string that has two modes excited and an outgoing string that is a tachyon. This was given in (\ref{320}),
\be \label{51}
\mA =\zeta_n {\cdot} \mV_n \zeta_m {\cdot} \mV_m + \zeta_n{\cdot} \zeta_m \mW_{n,m}~,
\ee
where it is important to remember that $\zeta_{n, \mu} \equiv \lam_{n, \mu}  - (\lam_n {\cdot} p) q_{\mu}$. 
We square the amplitude, 
\be \label{52}
|\mA|^2=  |\zeta_n {\cdot} \mV_n \zeta_m {\cdot}\mV_m|^2+  \( ( \zeta_n {\cdot} \mV_n \zeta_m{ \cdot} \mV_m)( \zeta_n{\cdot}\zeta_m \mW_{n,m})^* + \text{c.c.}\)+ |\zeta_n{\cdot }\zeta_m \mW_{n,m}|^2~,
\ee
 and sum over the polarizations by replacing $\sum_{\lam}(\lam_n^{\mu})^*\lam_n^{ \nu}\rightarrow \eta^{\mu \nu}$. We do this for each of the three terms above. For the first term we find that,
\be
  \sum_{\text{polarizations}}   |\zeta_n {\cdot} \mV_n \zeta_m {\cdot}\mV_m|^2 =(L_{\mu \nu} \mV_n^{\mu} \mV_n^{\nu})(L_{\al \beta}\mV_m^{\al}  \mV_m^{\beta})~,
  \ee
where we defined,
\be 
L_{\mu\nu}= \eta_{\mu \nu} - p_{\mu}q_{\nu} -p_{\nu}q_{\mu} + p^2 q_{\mu}q_{\nu}~,
\ee
which serves as a projection operator. $L_{\mu \nu}$ has the following properties, 
 \be
 L_{\mu \nu}p^{\nu}  = L_{\mu \nu}q^{\nu}=0 ~, \ \ \ \ \ L_{\al \beta} L_{\rho \sigma} \eta^{\beta \sigma} = L_{\al \rho} \ \ \ \ \ L_{\mu\nu}=L_{\nu \mu}~, \ \ \ \ L_{\mu \nu} \eta^{\mu \nu}=L_{\mu \nu}L^{\mu \nu} =D{-}2~.
 \ee 
Likewise,  the sum over polarizations of the second term in (\ref{52}) is, 
\be
\sum_{\text{polarizations}}  ( \zeta_n {\cdot} \mV_n {\zeta_m}{ \cdot} \mV_m)( \zeta_n{\cdot} \zeta_m )^*= L_{\mu \nu}\mV_n^{\mu}\mV_m^{\nu}~,
\ee
and finally, for the last term in (\ref{52}) we have,
\be
\sum_{\text{polarizations}} |\zeta_n{\cdot} \zeta_m|^2=D{-}2~,
 \ee
 where $D$ is the spacetime dimension (it arises from $\eta_{\mu \nu}\eta^{\mu\nu} = D{-}2$). Summing these three contributions, we can write them in the following instructive form, 
  \be \label{58}
 \sum_{\text{polarizations}} |\mA|^2 = L_{\mu \nu} L_{\al \beta}( \mV_n^{\mu}\mV_m^{\al} + \eta^{\mu \al} \mW_{n,m})( \mV_n^{\nu}\mV_m^{\beta} + \eta^{\nu \beta} \mW_{n,m})=  L_{\mu \nu} L_{\al \beta} \mA^{\mu \al} \mA^{\nu \beta}~.
 \ee
 This reproduces the answer (\ref{417}) that was claimed in the main body of the text. 
 
 Repeating this derivation for a few more examples, it becomes clear that in general the result is the equation (\ref{424}) given in the main body of the text. 

\subsection{Photon emission}\label{apd}
In the bulk of the paper we computed the amplitude of an excited string to decay into another excited string, via emission of a tachyon. In this appendix we modify the formulas for the case in which, instead of a tachyon, a photon is emitted. 

The three-point amplitude with an excited string of momentum $p$, another excited string of momentum $p'$, and a photon of polarization $\varepsilon$ and momentum $k$ is given by,
 \be
 \mA = \la V(z)\, V'(z') \,\varepsilon{\cdot}\partial X(w)  \,e^{i k{ \cdot }X(w) }\ra~,
 \ee
 where $V(z)$ and $V'(z')$ are the vertex operators for the excited strings. The amplitude is similar to the earlier amplitude we studied, (\ref{B1}), in which a tachyon is emitted -- the distinction is that the photon vertex operator has an additional $\varepsilon{\cdot}\partial X(w)$ factor. 
 
 Inserting the vertex operators for the excited strings, and performing the Wick contractions, the amplitude can be written as, 
 \be
 \mA = \prod_{n,a} \zeta_n^{(a)}{\cdot} \frac{\d}{\d {J}_n^{(a)}} \, {\zeta_n'}^{(a)}{\cdot} \frac{\d}{\d {J'_n}^{(a)}} \mA_{\text{gen}}^{\text{ph}}\Big|_{J = J'=0}~,
 \ee
 where the generating function is, 
 \be  \label{421photon}
\mA_{\text{gen}}^{\text{ph}}\! =\left(\varepsilon{\cdot}p' - \sum_{n,a} {nq{\cdot}p'{+}n\over 1{+}n q{\cdot}p'}q'{\cdot}{\cal V}_{n}\, J_{n}^{(a)}{\cdot}\varepsilon+ \sum_{n,a} {nq'{\cdot}p{+}n\over 1{+}nq'{\cdot}p}q{\cdot}{{\cal V}'}_{n}\, {J_{n}'}^{(a)}{\cdot}\varepsilon\right) \, {\cal A}_{\text{gen}}~,
\ee
where ${\cal A}_{\text{gen}}$ is the generating function (\ref{421}) for the case in which a tachyon is emitted. The distinction between the amplitude for photon emission versus tachyon emission is due to the  additional terms, linear in the photon polarization, which are generated by  all the possible Wick contractions of the operator $\varepsilon{\cdot}\partial X$. In fact, a simple way of getting (\ref{421photon}) is to generalize (\ref{421}) to the case in which all three strings are excited, and then specialize to the case in which one of the excited strings is in the first excited state.

Turning to the kinematics, in the frame in which an excited string  at rest decays,  it is given by,
 \bea \nn
 p&=&\sqrt{2N{-}2}(1,\vec{0})\\ \nn
 p'&=&-(E',\omega \sin\theta,\omega \cos\theta,\vec{0})\\
 k&=&-\omega(1,-\sin\theta, -\cos\theta,\vec{0})~,
 \eea
with
\be
\omega={N-N'\over \sqrt{2N-2}}\,,\quad E'={N+N'-2\over \sqrt{2N-2}}~.
\ee
This kinematics is a slight variation of the kinematics in the main body (\ref{35}) for tachyon emission.

{\setstretch{1}

\bibliographystyle{utphys}

\end{document}